\documentclass[twopage,11pt]{article}
\usepackage{amsmath}
\usepackage{graphicx}
\usepackage{amsfonts}

\numberwithin{equation}{section}

\title{Wormhole solutions in 5D Kaluza-Klein theory as string-like 
objects}

\author{Vladimir Dzhunushaliev
\thanks{e-mail: dzhun@hotmail.kg}\\
\textit{Dept. Phys. and Microel. Engineer.,} \\
\textit{Kyrgyz-Russian Slavic University}\\
\textit{Bishkek, Kievskaya Str. 44, 720021, Kyrgyz Republic}}

\date{}

\begin{document}
\maketitle

\date{}

\maketitle

\pagestyle{myheadings} 
\thispagestyle{plain}         
\markboth{V. Dzhunushaliev, D. Singleton and T. Nikulicheva}
{Nonperturbative calculational method in quantum field theory} 
\setcounter{page}{1}  

\begin{abstract}
The detailed numerical and analytical approximate analysis of wormhole-like 
solutions in 5D Kaluza-Klein gravity is given. It is shown that some part 
of these solutions with $E \approx H, E>H$ relation between electric $E$ and magnetic $H$ 
fields  can be considered as a superthin and superlong 
gravitational flux tube filled with electric and magnetic fields, namely 
$\Delta-$strings. The solution behaviour near hypersurface 
$ds^2=0$ and the model of electric charge on the basis of $\Delta-$string 
are discussed. The comparison of the properties of $\Delta-$string and ordinary 
string in string theory is carried out. Some arguments are given that fermionic 
degrees of freedom can be build in on the $\Delta-$string. These degrees of freedom 
are connected with quantum wormholes of spacetime foam. It is shown that the 
natural theory for these spinor fields is supergravity. 
\end{abstract}

\section{Introduction}
\label{introduction}

Up to now the inner structure of elementary particles (such as electron) is 
unknown. There are two conceptions of their structue in physics: the first one 
is Newton's conception and the second one is Einstein's. The essence of the first 
approach is that electron is the structurless pount-like object with mass $m$, 
charge $e$ and spin $\hbar /2$. There are many well known problems in this approach, for example, 
the infiniteness of own electromagnetic energy, loop divergences in quantum 
electrodynamics and so on. Now this point of view is extended in one dimension 
and is known as string theory. In string theory elementary particles are strings with 
vibrations which numerate elementary particles. On strings live fermion degrees of freedom 
which are the origin of spin. There are many attractive characteristic properties of 
string-like conception of elementary particles. They are well known and we will 
not discuss here. But there is the fundamental insoluble problem: what kind of matter 
is inside string ? Evidently that we \textit{must} have some theory for string matter. 
In this theory we will have another ''elementary particles`` and so on up to infinfity. 
This is the big problem for string theory.
\par 
Another point of view on the inner structure of elementary particles is presented by 
Einstein. His point of view is that electron has a complicated inner structure 
with a core filled with fields and where gravity plays a fundamental role. As 
an example of such conception a wormhole can be considered. In such picture 
one part of the wormhole entraps electric force lines and another one exhausts 
these force lines. For an observer at the infinity the first part of wormhole 
is like to (-) charge and another one (+) charge. This model of electric charge 
has many problems such as: the absence of spin $\hbar /2$, nonquantized electric 
charge and so on. One of the problem in this approach is that for the wormhole solutions in 4D 
Einstein gravity an exotic matter is necessary, whereas Einstein's idea 
is that the wormhole \textit{must be a vacuum solution} filled with a 
fundamental field (for example, electric field). 
\par 
In this letter we would like to present and discuss a hybrid between wormhole and string. 
It is vacuum solutions in 5D Kaluza-Klein gravity. In fact some part of these 
solutions can be considered as a superlong and superthin gravitational flux tube 
filled with electric and magnetic field. The thickness of this object is so 
small that near to the point of attachment to an external Universe quantum wormholes  
of a spacetime foam will appear between this object and the Universe. 
This is like a delta of the river flowing into the sea. 
We call such objects as a $\Delta-$string. The cross section of the tube is an arbitrary 
parameter and can be choosen in Planck region consequently it is a superthin 
tube. The length of the tube can be arbitrary long and depends on the relation 
between electric $E$ and magnetic $H$ fields. It can be infinite if $E = H$. 
The electromagnetic potential $A_\mu$ is connected with off-diagonal components 
$G_{5\mu}$ of 5D metric. 

\section{Wormhole-like solutions in 5D Kaluza-Klein gravity}
\label{wormhole}

In this section we would like to present wormhole-like solutions in 5D Kaluza-Klein 
gravity. In the initial Kaluza-Klein interpretation 5D metric does not depend 
on the fifth coordinate. There are two reasons for this statement: 
\begin{enumerate}
	\item 
	$5^{th}-$dimension has a Planck length. One of the paradigm of quantum gravity is that 
	the Planck length is the minimal length in the nature and we can not say anything 
	about spacetime structure inside Planck length. In such case any physical fields 
	can not be variable inside Planck cell. Consequently if $5^{th}-$dimension is in 
	the Planck region then physical degrees of freedom have not dependence on 
	$5^{th}-$coordinate.
	\item 
	5D spacetime is a total space of U(1) principal bundle, $5^{th}-$dimension 
	is a fibre of the principal bundle and topologically it is the U(1) structure group. 
	It is well known that any group $G$ is a symmetrical space and the metric of 
	$G$ must be symmetrical and coincides with a Killing metric up to a constant 
	$\phi$ which can depend on the coordinates of the base of principal bundle. 
\end{enumerate}
Thus these two items give us physical reasonings that any physical fields in 5D 
kaluza-Klein gravity depend only on the spacetime coordinates $x^\mu, \mu=0,1,2,3$.
\par
In common case the 5D metric $G_{AB}$ ($A,B = 0,1,2,3,5$) has the 
following form 
\begin{equation}
	ds^2 = g_{\mu \nu}dx^\mu dx^\nu - \phi^2 \left( dx^5 + A_\mu dx^\mu \right)^2 
\label{sec1-1a}
\end{equation}
$\mu , \nu = 0,1,2,3$ are the 4D indices. According to the Kaluza - Klein 
point of view $g_{\mu \nu}$ is the 4D metric; 
$A_\mu$ is the usual electromagnetic potential and $\phi$ is some scalar 
field. The 5D Einstein vacuum equations are 
\begin{equation}
R_{AB} - \frac{1}{2}G_{AB} R = 0.
\label{sec1-2}
\end{equation}
Now we would like to present the spherically symmetric solutions with
nonzero flux of electric and magnetic fields \cite{dzhsin1}.
For our spherically symmetric 5D metric we take
\begin{eqnarray}
ds^2 & = & \frac{dt^{2}}{\Delta(r)} - l_0^2 \Delta(r) e^{2\psi (r)}
\left [d\chi +  \omega (r)dt + Q \cos \theta d\varphi \right ]^2
\nonumber \\
&-& dr^{2} - a(r)(d\theta ^{2} +
\sin ^{2}\theta  d\varphi ^2),
\label{sec1-1}
\end{eqnarray}
where $\chi $ is the 5$^{th}$ extra coordinate;
$r,\theta ,\varphi$ are $3D$  spherical-polar coordinates;
$r \in \{ -r_H , +r_H \}$
($r_H$ may be equal to $\infty$), $l_0$ is some constant; $Q$ is 
the magnetic charge as $(\theta, \varphi)$ -component of the Maxwell tensor 
$F_{\theta, \varphi} = -Q \sin\theta$.
\par 
The metric \eqref{sec1-1} gives us the following components of electromagnetic potential 
$A_\mu$ 
\begin{equation}
A_0 = \omega (r) \quad 
\text{and} \quad 
A_\varphi = Q \cos \theta
\label{sec1-2e}
\end{equation}
that gives the Maxwell tensor 
\begin{equation}
F_{10} = \omega (r) ' \quad 
\text{and} \quad 
F_{23} = -Q \sin \theta .
\label{sec1-2f}
\end{equation}
This means that we have radial Kaluza-Klein 
electric $E_r \propto F_{01}$ and magnetic 
$H_r \propto F_{23}$ fields.
\par
Substituting this ansatz into the 5D Einstein vacuum equations
\eqref{sec1-2} gives us 
\begin{eqnarray}
\frac{\Delta ''}{\Delta} - \frac{{\Delta '}^2}{\Delta^2} + 
\frac{\Delta 'a'}{\Delta a} + \frac{\Delta ' \psi '}{\Delta} + 
\frac{q^2}{a^2 \Delta ^2}e^{-4 \psi} & = & 0,
\label{sec1-3}\\
\frac{a''}{a} + \frac{a'\psi '}{a} - \frac{2}{a} +
\frac{Q^2}{a^2} \Delta e^{2\psi} & = & 0,
\label{sec1-5}\\
\psi '' + {\psi '}^2 + \frac{a'\psi '}{a} -
\frac{Q^2}{2a^2} \Delta e^{2\psi} & = & 0,
\label{sec1-6}\\
- \frac{{\Delta '}^2}{\Delta^2} + \frac{{a'}^2}{a^2} - 
2 \frac{\Delta ' \psi '}{\Delta} - \frac{4}{a} + 
4 \frac{a' \psi '}{a} + 
\frac{q^2}{a^2 \Delta ^2} e^{-4 \psi} + 
\frac{Q^2}{a^2} \Delta e^{2\psi} & = & 0
\label{sec1-7}
\end{eqnarray}
$q$ is electric charge; these equations are derived after substitution the expression 
\eqref{sec1-7e} for the electric field in the initial Einstein's equatons 
\eqref{sec1-3}-\eqref{sec1-7}. The 5D $(\chi t)$-Einstein's equation (4D Maxwell 
equation) is taken as having the following solution 
\begin{equation}
  \omega ' = \frac{q}{l_0 a \Delta ^2} e^{-3 \psi} .
\label{sec1-7e}
\end{equation}
For the determination of the physical sense of the constant $q$ let us 
write the $(\chi t)$-Einstein's equation in the following way :
\begin{equation}
\left( l_0 \omega ' \Delta ^2 e^{3\psi} 4 \pi a \right)' = 0.
\label{sec1-7a}
\end{equation}
The 5D Kaluza - Klein gravity after the dimensional reduction says us that 
the Maxwell tensor is 
\begin{equation}
F_{\mu \nu} = \partial_\mu A_\nu - \partial _\nu A_\mu . 
\label{sec1-7e1}
\end{equation}
That allows us to write in our case the electric field 
as $E_r = \omega '$. 
Eq.\eqref{sec1-7a} with the electric field defined by \eqref{sec1-7e1} 
can be compared with the Maxwell's equations in a continuous medium 
\begin{equation}
\mathrm{div} \mathcal {\vec D} = 0
\label{sec1-7e2}
\end{equation}
where $\mathcal {\vec D} = \epsilon \vec E$ is an electric displacement 
and $\epsilon$ is a dielectric permeability. Comparing Eq. \eqref{sec1-7a}
with Eq. \eqref{sec1-7e2} we see that the magnitude 
$q/a = \omega ' \Delta^2 e^{3\psi}$ is like to the electric displacement 
and the dielectric permeability is $\epsilon = \Delta^2 e^{3\psi}$. 
It means that $q$ can be taken as the Kaluza-Klein electric charge 
because the flux of electric field is 
$\mathbf{\Phi} = 4\pi a\mathcal D = 4\pi q$.
\par 
For $r=0$ Eq. \eqref{sec1-7} gives us the
following relationship between the Kaluza-Klein electric $q$ 
and magnetic $Q$ charges
\begin{equation}
\label{sec1-7d}
1 = \frac{q^2 + Q^2}{4a(0)}
\end{equation}
where $a(0) = a(r=0)$. 
\par
As the relative strengths of the Kaluza-Klein fields are
varied it was found \cite{dzhsin1} that the solutions to the metric in
Eq. \eqref{sec1-1} evolve in the following way :
\begin{enumerate} 
\item 
$0 \leq Q < q$. The solution is \emph{a regular flux tube}. 
The throat between the surfaces at $\pm r_H$ is filled with both 
electric and magnetic fields. The longitudinal
distance between the $\pm r_H$ surfaces increases, and
the cross-sectional size does not increase as rapidly
as $r \rightarrow r_H$ with $q \rightarrow Q$. Essentially,
as the magnetic charge is increased one can think that
the $\pm r_H$ surfaces are taken to $\pm \infty$ and 
the cross section becomes constant. The radius $r= \pm r_H$ is defined by the 
following way $ds^2(r = \pm r_H) = 0, \Delta(\pm r_H)=0$. 
\item 
$Q = q$. In this case the solution is \emph{an infinite flux tube} filled
with constant electric and magnetic fields. The cross-sectional
size of this solution is constant ($ a= const.$). 
\item 
$0 \leq q < Q$. In this case we have \emph{a singular flux tube} located 
between two (+) and (-) electric and magnetic 
charges located at $\pm r_{sing}$. Thus the longitudinal 
size of this object is again finite, but now the cross 
sectional size decreases as $r \rightarrow r_{sing}$. At 
$r = \pm r_{sing}$ this solution has real singularities which 
we interpret as the locations of the charges. 
\end{enumerate} 
\par
\begin{figure}
\begin{center}
\fbox{
\includegraphics[height=5cm,width=7cm]{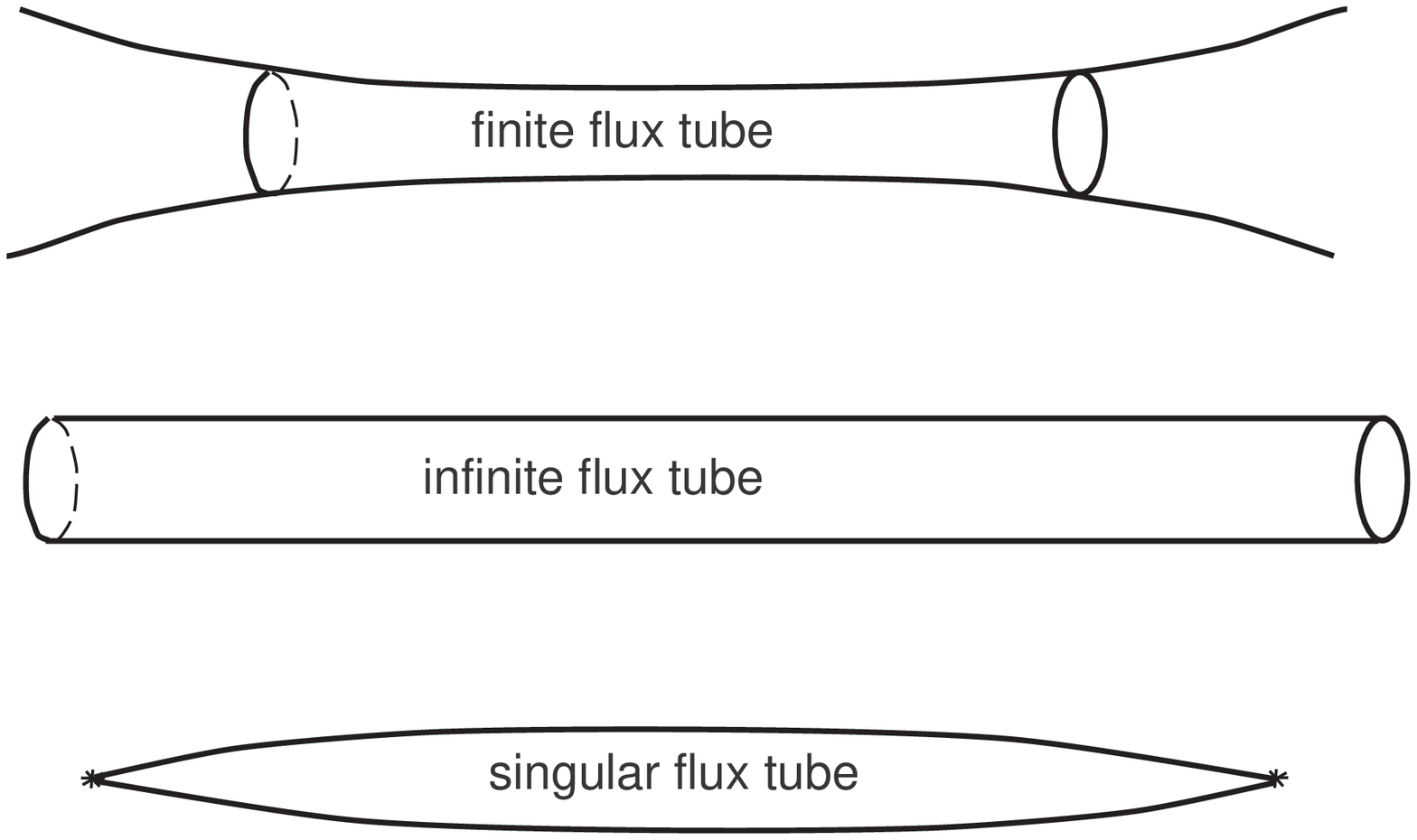}}
\caption{The gravitational flux tubes with the different relation between 
electric $q$ and magnetic $Q$ charges.}
\label{fig1}
\end{center}
\end{figure}

 
Schematically the evolution of the solution from a regular flux tube, to
an infinite flux tube, to a singular flux tube, as the
relative magnitude of the charges is varied, is presented in
Fig.\eqref{fig1}. 
\par
Most important for us here is the case with $\delta = 1 - \frac{Q}{q} \ll 1$. 
From Eq. \eqref{sec1-7d} we have 
$q = 2 a_0 \sin (\pi /4 + \delta) \approx a_0 \sqrt{2} (1 + \delta)$ 
and $Q \approx a_0 \sqrt{2} (1 - \delta)$. 

\section{Infinite flux tube solution}
\label{infinite}

In this section we would like to present in details the infinite flux tube solution 
with $E =H$ as this metric is very good approximation for central part of solutions 
with $\delta \ll 1$. The metric is the solution of Eq's \eqref{sec1-3}-\eqref{sec1-7}
\begin{eqnarray}
	ds^2 &=& \cosh \left( \frac{r\sqrt{2}}{q} \right) dt^2 - 
	\left[ 
		d\chi + \sqrt{2} \sinh \left( \frac{r\sqrt{2}}{q} \right)dt + 
		Q \cos \theta d\varphi
	\right]^2 - 
\nonumber \\
	&&dr^2 - \frac{q^2}{2}
	\left(
		d \theta ^2 + \sin^2 \theta d\varphi^2
	\right) = 
\label{inf-10}\\
	&&\left[
		1 - \sinh^2 \left( \frac{r\sqrt{2}}{q} \right)
	\right] dt^2 - 
	Q^2 \left( \frac{d\chi}{Q} + \cos \theta d \varphi \right)^2 - \frac{q^2}{2}
	\left(
		d \theta ^2 + \sin^2 \theta d\varphi^2
	\right) - 
\nonumber \\
  && dr^2 - 2\sqrt{2} \sinh \left( \frac{r\sqrt{2}}{q} \right)
  \left( d\chi + Q \cos \theta d \varphi \right) dt .
\label{inf-20}
\end{eqnarray}
From the last form \eqref{inf-20} of the metric we see that the terms 
\begin{equation}
	dl^2 = Q^2 \left( \frac{d\chi}{Q} + \cos \theta d \varphi \right)^2 + \frac{q^2}{2}
	\left(
		d \theta ^2 + \sin^2 \theta d\varphi^2
	\right)
\label{inf-30}
\end{equation}
are like to the standard metric 
\begin{equation}
	dl^2 = \left( \frac{d\chi}{Q} + \cos \theta d \varphi \right)^2 + 
	\left(
		d \theta ^2 + \sin^2 \theta d\varphi^2
	\right)
\label{inf-40}
\end{equation}
on the $S^3-$sphere where $S^3-$sphere is presented as the Hopf bundle $S^3 \rightarrow S^2$ 
with the $S^1-$sphere as the fibre. The first term of Eq. \eqref{inf-20} tells us that the time $t$ 
is not everywhere the true time coordinate as for 
$r > r_1$ \footnote{the parameter $r_1$ is defined as follows: 
$\sinh^2 \left( \frac{r_1\sqrt{2}}{q} \right) = 1$} the metric component 
$G_{tt} = 1 - \sinh^2 \left( \frac{r\sqrt{2}}{q} \right) < 0$. 
\par 
The metric \eqref{inf-10} is very interesting in the following context. It has the form 
\textit{of dynamical splitting off the $5^{th}-$dimension as $G_{55} = 1$.} The origin 
of such splitting off is \textit{the presence of off-diagonal components of 5D metric 
$G_{5t}$ and $G_{5}\varphi$} which are electromagnetic potential components. From the 
physical point of view it is the consequence of the presence of electric and magnetic 
fields with very special constraint: $E=H$. Mathematically it is an exceptional 
solution of \eqref{sec1-3}-\eqref{sec1-7} equations. 
\par 
Another interesting question from the infinite flux tube metric \eqref{inf-10} is the 
question about clock reading. The metric \eqref{inf-20} shows that clock reading is 
\begin{equation}
	d\tau_5 =  dt \sqrt{1 - \sinh^2 \left( \frac{r \sqrt{2}}{q} \right)}= 
	dt\sqrt{G_{00} - A_0^2}.
\label{inf-50}
\end{equation}
Let us compare this 5D infinite flux tube solution with the Levi-Civita - Robertson - Bertotti 
flux tubes solutions \cite{Levi-Civita}. It is an infinite flux tube filled 
with parallel electric $F_{01}$ and and magnetic $F_{23}$ fields 
\begin{eqnarray} 
ds^2 &=& a^2\left (\cosh^2 \zeta dt^2 - d\zeta^2 - d\theta ^2 -
\sin ^2 \theta d\varphi ^2\right ),
\label{inf-60} \\ 
F_{01} &=& \rho ^{1/2} \cos\alpha, \;\;\; \; \; \; \;
F_{23} = \rho ^{1/2}\sin\alpha, 
\label{inf-65} 
\end{eqnarray} 
where 
\begin{equation} 
G^{1/2} a \rho ^{1/2} = 1. 
\label{inf-80} 
\end{equation} 
$\alpha$ is an arbitrary constant angle; $a$ and $\rho$ are constants defined 
by Eq. (\ref{inf-80}); $a$ determines the cross section of the tube and $\rho$ 
the magnitude of the electric and magnetic fields; 
$G$ is Newton's constant ($c=1$, the speed of light); 
$F_{\mu\nu}$ is the electromagnetic field tensor. For $\cos\alpha = 1$ 
($\sin\alpha = 1$) one has a purely electric (or magnetic) field. 
This 4D solution is the analog of the 5D flux tube solution with only one 
difference: electric and magnatic fields can be different. Clock reading for 4D solution 
is 
\begin{equation}
	d\tau_4 = \cosh \left( \frac{r \sqrt{2}}{q} \right)dt .
\label{inf-90}
\end{equation}
Physically this remark allows us to determine experimentally: is the dimension 
of our spacetime four or more. 

\section{Superlong solutions}
\label{superlong}

Now we would like to consider the case with $\delta = 1 - Q/q \ll 1$. 
In this section we will show that \emph{every such object with cross 
section $\approx l^2_{Pl}$ can be considered as a string-like object}. 
\par
\begin{figure}
\begin{center}
\fbox{
\includegraphics[height=5cm,width=5cm]{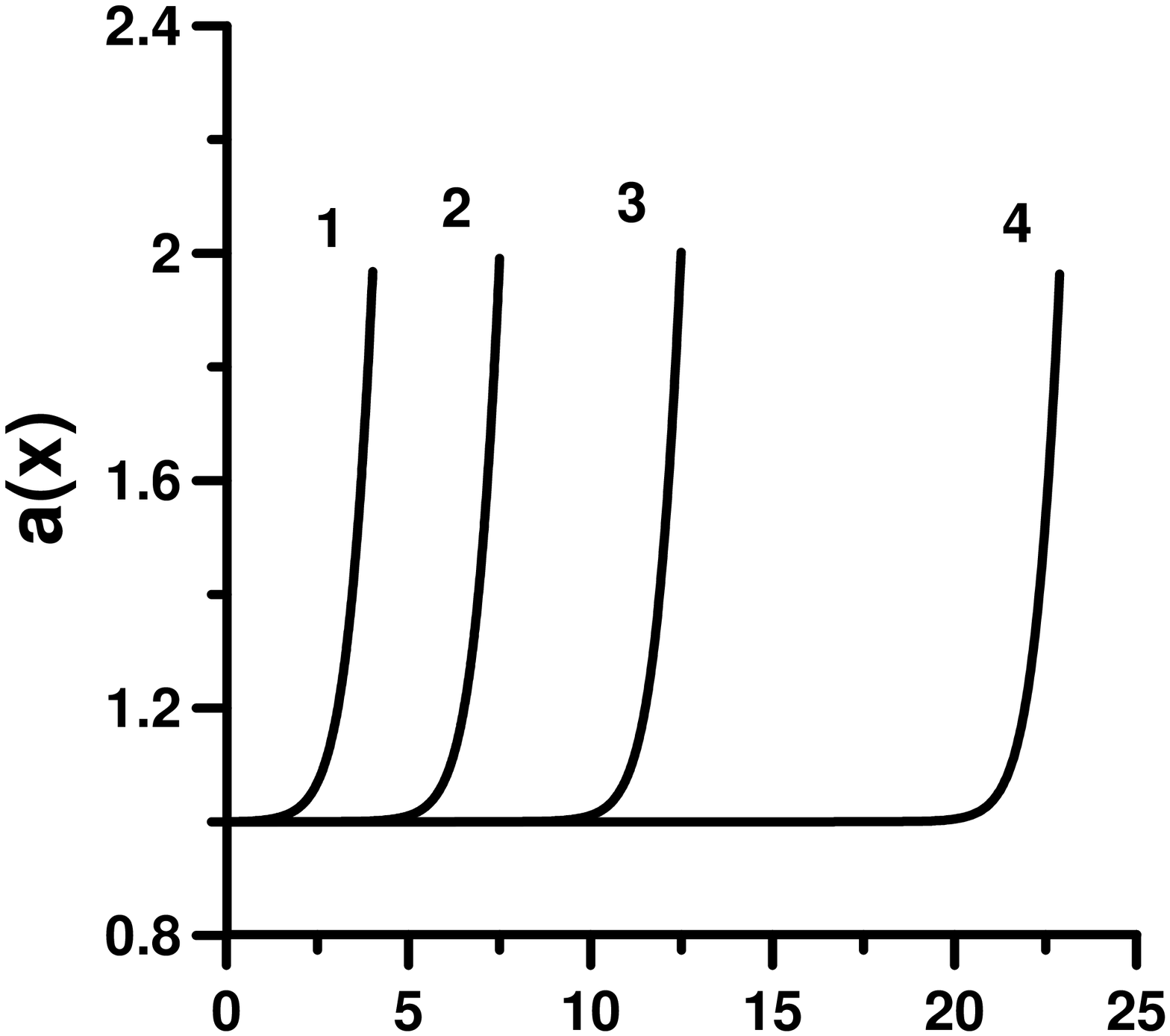}}
\caption{The functions $a(x)$ in accordance with the different relation between electric and 
magnetic charges. The curves 1,2,3,4 correspond to the following magnitudes of 
$\delta = 10^{-3}, 10^{-6}, 10^{-10}, 10^{-13}$.}
\label{fig2}
\fbox{
\includegraphics[height=5cm,width=5cm]{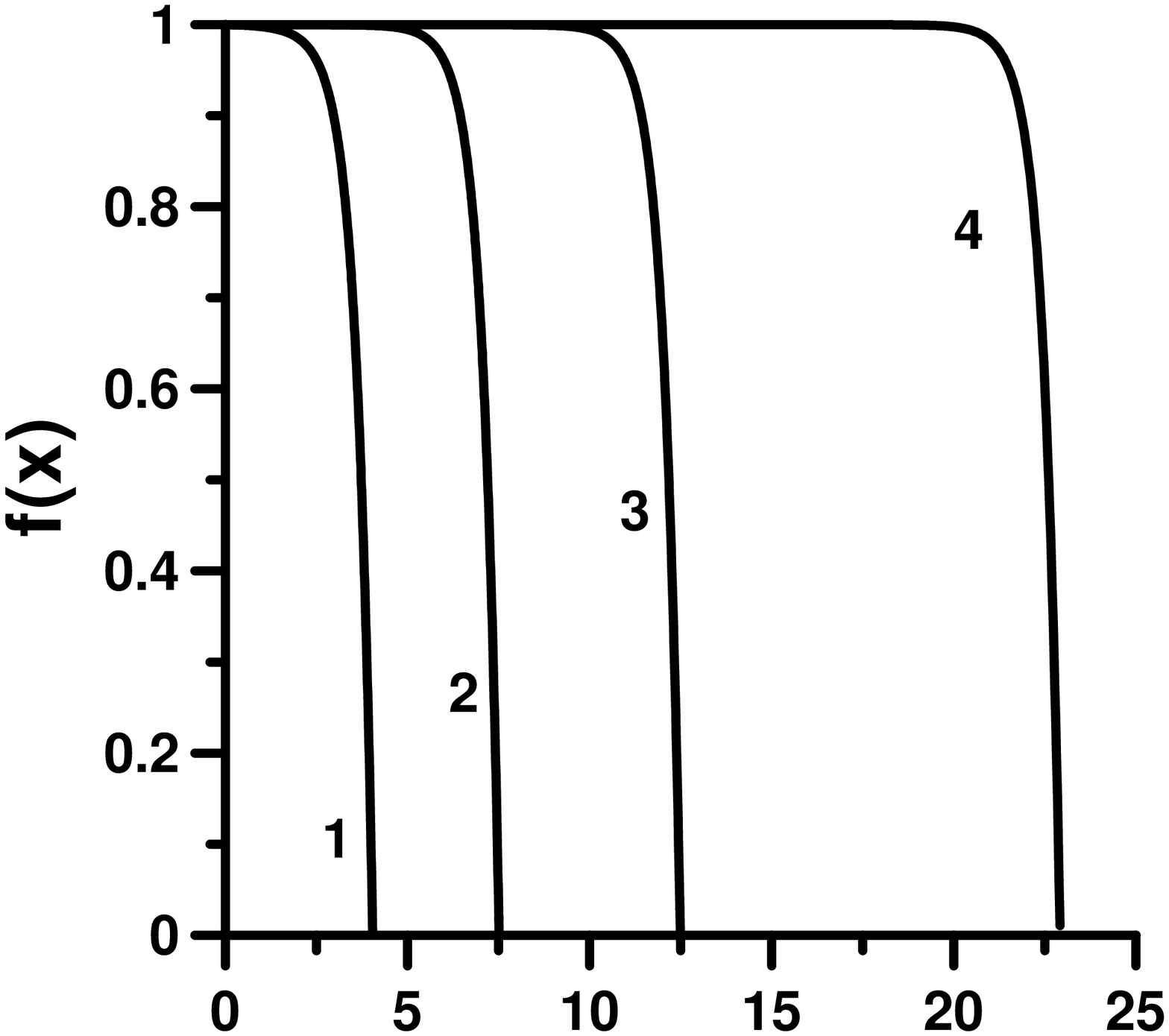}}
\caption{The functions $f(x)$ in accordance with the different relation between electric and 
magnetic charges. The curves 1,2,3,4 correspond to the following magnitudes of 
$\delta = 10^{-3}, 10^{-6}, 10^{-10}, 10^{-13}$.}
\label{fig3}

\fbox{
\includegraphics[height=5cm,width=5cm]{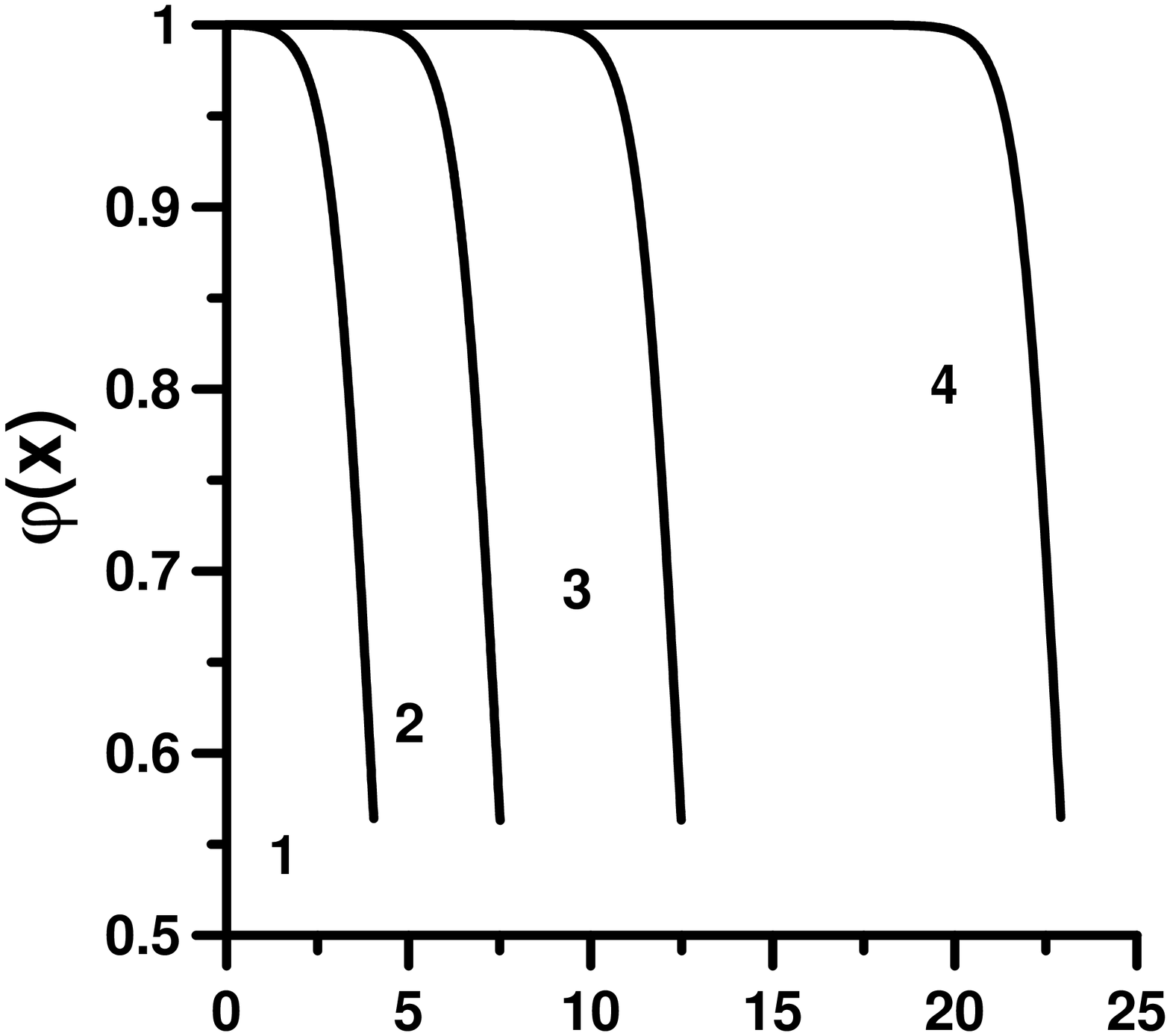}}
\caption{The functions $\phi(x)$ in accordance with the different relation between electric and 
magnetic charges. The curves 1,2,3,4 correspond to the following magnitudes of 
$\delta = 10^{-3}, 10^{-6}, 10^{-10}, 10^{-13}$.}
\label{fig4}
\end{center}
\end{figure}

 
At first we will investigate solutions of Eq's 
\eqref{sec1-3}-\eqref{sec1-7} for different $\delta$'s. 
The numerical investigations of these equations set show us that the solution 
is very insensitive to $\delta \ll 1$. In order to avoid this 
problem we introduce the following new functions 
\begin{eqnarray}
  \Delta (x) & = & \frac{f(x)}{\cosh^2 (x)}, 
\label{sec4-10a}\\
  \psi (x) & = & \phi (x) + \ln \cosh (x) 
\label{sec4-10b}
\end{eqnarray}
where $x = r/a(0)^{1/2}$ is a dimensionless radius. In the case $f(x) = 1$ and $\phi (x) = 0$ 
we have the above-mentioned infinite flux tube. Thus we have the following 
equations for the functions \eqref{sec4-10a} \eqref{sec4-10b}
\begin{eqnarray}
  f'' - 4 f' \tanh (x) -2 f \left( 1 - 3 \tanh^2(x) \right) - 
  \frac{\left( f' - 2 f \tanh (x) \right)^2}{f} + &&  
\nonumber \\   
  \frac{a'}{a} \left( f' - 2 f \tanh (x) \right) + 
  \left( f' - 2 f \tanh (x) \right) 
  \left( \phi ' + \tanh (x) \right) + 
  \frac{q^2}{a^2 f} e^{-4 \phi}  = 0 , && 
\label{sec4-20a}\\
  a'' - 2 + a' \left( \phi ' + \tanh (x) \right) + 
  \frac{Q^2 f}{a} e^{2 \phi} = 0 , && 
\label{sec4-20b}\\
  \phi '' + \frac{1}{\cosh ^2(x)} + \left( \phi ' + \tanh (x) \right)^2 + 
  \frac{a'}{a}\left( \phi ' + \tanh (x) \right) - 
  \frac{Q^2 f}{2a^2} e^{2 \phi} = 0 && .
\label{sec4-20c}  
\end{eqnarray}
The corresponding solutions of these equations with the different $\delta$ 
are presented on Fig's \eqref{fig2}-\eqref{fig4}.
We see that (at least in the investigated area of $\delta$) at the point 
$r = \pm r_H$  the function $a(r_H) \approx 2 a_0$ (here $r_H$ is the point 
where $\Delta (r_H) = 0$). The most interesting aspect
is that $a(r_H)$ does not grow as $r \rightarrow \pm r_H$ and consequently 
\emph{the flux tube 
with $\delta \ll 1$ (in the region $|r| \leq r_H$) can be considered 
as a string-like object attached to two Universes.}
\par 
All this shows us that the conditions which are necessary
for the consideration of the presented solution as a string-like object definitely 
are satysfied in the region $|r| \leq r_H$ but outside of this region 
the cross section will increase and spacetime is asymptotically flat. 
It is easy to see for the special case $Q = 0$ \cite{dzhsin1}. 
In this case there is the exact solution 
\begin{eqnarray}
a & = & r^{2}_{0} + r^{2},
\label{sec3-9a}\\
\Delta & = & \frac{q}{2r_{0} q}
\frac{r^{2}_{0} - r^{2}}{r^{2}_{0} + r^{2}},
\label{sec3-9b}\\
\psi &=& 0 ,
\label{sec-9c}\\
\omega & = & \frac{4r_{0}}{q}
\frac{r}{r^{2}_{0} - r^{2}}
\label{sec3-9d}
\end{eqnarray}
and we see that $a \approx r^2$ 
(at $a \rightarrow \infty$) and $g_{tt} \approx -1 + \mathcal O (1/r^2)$, 
the same is for the others cases with $Q < q$. 
\par
\begin{figure}
\begin{center}
\fbox{
\includegraphics[height=5cm,width=9cm]{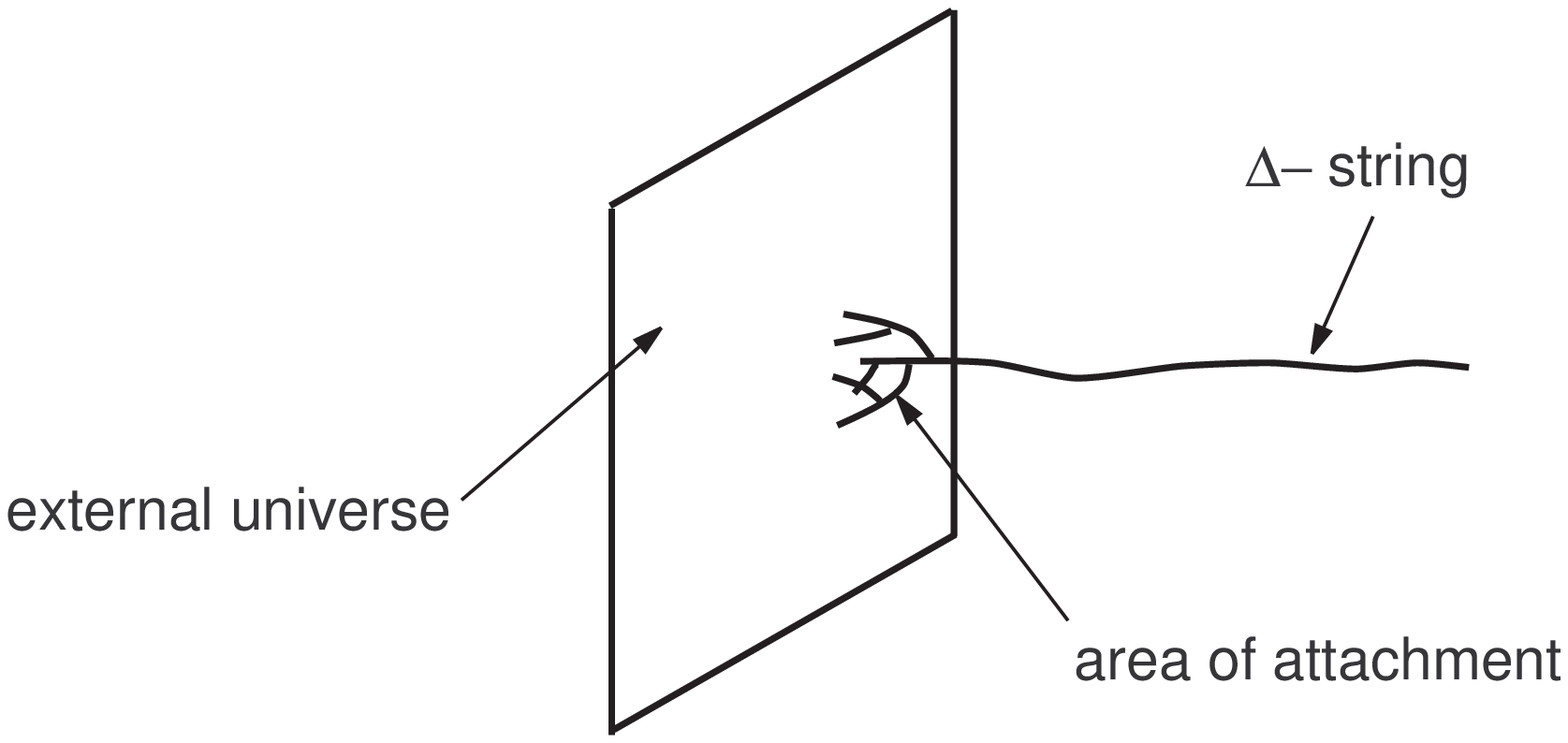}}
\caption{The attachment point of $\Delta-$string to an external Universe.}
\label{delta}
\end{center}
\end{figure}

The thickness of the gravitational flux tube 
can be arbitrary small and we choose its in the Planck region. In this case 
near to the attachment point of the tube to an external Universe the quantum 
wormholes of a spacetime foam will appear between this object and the Universe. 
This is like a delta of the river flowing into the sea, see Fig. \ref{delta}. 
This remark allows us to call the super-long and thin gravitational flux tube 
as the $\Delta-$string. 

\section{The metric close to $r=\pm r_H$}
\label{horizon}

Now we would like to show that the $\Delta-$string solution is nonsingular 
at the points $\pm r_H$ where $\Delta(\pm r_H) = 0$. For this we 
investigate the solution behaviour near to the point $|r| \approx r_H$ where 
\begin{eqnarray}
  a(r) &=& a_0 + a_1 \left ( r-r_H \right ) + 
  a_2 \left ( r-r_H \right )^2 + \cdots ,
    \label{sec1:51}\\
    \psi(r) & = & \psi_H + \psi_1 \left ( r-r_H \right ) + 
    \psi_2 \left ( r-r_H \right )^2 + \cdots ,
    \label{sec1:52}\\
    \Delta(r) & = & \Delta_1 \left( r - r_H \right) + 
    \Delta_2 \left( r - r_H \right)^2 + \cdots .
    \label{sec1:54}
\end{eqnarray}
The substitution in Eq's \eqref{sec1-3}-\eqref{sec1-7} gives us the following 
solution
\begin{eqnarray}
    \Delta_1 & = & \pm q e^{-2\psi_H},
    \quad (+) \quad \text{for} \quad r \rightarrow -r_H 
  \quad \text{and} 
  \quad (-) \quad \text{for} \quad r \rightarrow +r_H ,
    \label{sec1:56}\\
    \psi_2 &=& -\psi_1 \frac{a_1 + a_0\psi_1}{2a_0},
    \label{sec1:57}\\   
    \Delta_2 &=& \frac{-3a_0 \psi_1 + a_1}{2 a_0},
    \label{sec1:58}\\   
    a_2 &=& \frac{2 - a_1 \psi_1}{2}.
    \label{sec1:59}
\end{eqnarray}
In this case equation \eqref{sec1-7e} has the following behaviour near to the 
points $r=\pm r_H$  
\begin{equation}
    \omega'(r) = \frac{a_0 e^{\psi_H}}{q} \frac{1}{\left( r - r_H \right)^2} + 
    \omega_1 + \mathcal{O}\left( r - r_H \right)
    \label{sec1:59a}
\end{equation}
where $\omega_1$ is some constant depending on $a_{0,1}, \psi_{1,2}, \Delta_{1,2}$. 
It leads to the following $\omega(r)$ behaviour 
\begin{equation}
    \omega(r) = -\frac{a_0 e^{\psi_H}}{q} \frac{1}{\left( r - r_H \right)} + 
    \omega_0 + \mathcal{O}\left( r - r_H \right)
    \label{sec1:59b}
\end{equation}
where $\omega_0$ is some integration constant. The $G_{tt}$ metric component is 
\begin{equation}
    G_{tt} = \frac{a(r)}{\Delta(r)} - 
    \frac{\Delta(r) e^{2 \psi(r)}}{a(r)} \omega^2(r) = 
    -e^{2\psi_H} \frac{2qe^{-\psi_H} \omega_0 - a_1 - a_0 \psi_1}{q} + 
    \mathcal{O} \left( r - r_H \right).
    \label{sec1:59c}
\end{equation}
Then the metric \eqref{sec1-1} has the following approximate behaviour 
near to $r = \pm r_H$ points \begin{equation}
\begin{split}
    ds^2 = &\left[ g_H +    \mathcal{O} \left( r - r_H \right) \right] dt^2 - 
    \mathcal{O} \left( r - r_H \right) 
    \left( d\chi + Q \cos\theta d \varphi \right)^2 - \\
    &\left[ e^{\psi_H} + \mathcal{O} \left( r - r_H \right) \right]
    dt \left( d\chi + Q \cos\theta d \varphi \right) - dr^2 - \\
    &\left[ a(r_H) + \mathcal{O} \left( r - r_H \right] \right) 
    \left( d\theta^2 + \sin^2 \theta d\varphi^2 \right) \approx \\
    &e^{\psi_H} dt \left( d\chi + Q \cos\theta d \varphi \right) - 
    dr^2 - a(r_H) \left( d\theta^2 + \sin^2 \theta d\varphi^2 \right)
\end{split} 
\label{sec1:59e}
\end{equation}
where 
$g_H = -e^{2\psi_H} \left (2qe^{-\psi_H} \omega_0 - a_1 - a_0 \psi_1 \right )/q$. 
It means that at the points $r = \pm r_H$ the metric \eqref{sec1-1} is nonsingular. 

\section{Approximate analytical solution close to pure electric solution 
with $Q=0$}
\label{approximate}

In this section we would like to investigate analytically the metric close to 
exact solution with $Q=0$. 
\begin{eqnarray}
  a(r) &=& a_0(r) + \delta a(r) ,
\label{sec3-10}\\
  \Delta(r) &=& \Delta_0(r) + \delta\Delta (r) ,
\label{sec3-20}\\
  \psi(r) &=& \psi_0(r) + \delta\psi(r) 
\label{sec3-30}
\end{eqnarray} 
and $Q$ is small parameter $Q^2/a(0) \ll 1$. 
\begin{eqnarray}
  a_0(r) &=& r^2 + r_0^2 ,
\label{sec3-40}\\
  \Delta_0(r) &=& r_0^2 - r^2 ,
\label{sec3-50}\\
  \psi_0(r) &=& 0 
\label{sec3-60}
\end{eqnarray} 
the functions $a_0(r), \Delta_0(r)$ and $\psi_0(r)$ are the solutions 
of Einstein equations with $Q=0$. The variation of equations 
\eqref{sec1-3}-\eqref{sec1-7} with respect to small perturbations 
$\delta a(r), \delta\Delta (r)$ and $\delta\psi(r)$ give us the following 
equations set 
\begin{eqnarray}
  \delta\psi + \frac{a'_0}{a_0}\delta\psi ' - 
  \frac{Q^2 \Delta_0}{2a_0^3} &=& 0 ,
\label{sec3-70}\\
  \delta a '' + a'_0 \delta\psi ' + \frac{Q^2 \Delta_0}{a_0^2} &=& 0 ,
\label{sec3-80}\\
  \delta\Delta '' - \frac{a'_0}{a_0}\delta\Delta'  - 
  \frac{\Delta'_0 \delta a'}{a_0} + 
  \frac{\Delta'_0 a'_0}{a_0^2} \delta a + 3 \Delta'_0 \delta\psi ' + 
  2\frac{\delta\Delta}{a_0} - 2\frac{\Delta_0}{a_0^2} \delta a &=& 0 .
\label{sec3-90}
\end{eqnarray} 
The solution of this system is 
\begin{eqnarray}
  \delta\psi(r) &=& -\frac{1}{4} \frac{Q^2}{r^2 + r_0^2} ,
\label{sec3-95}\\
  \delta a(r) &=& - \frac{Q^2}{2} \frac{r}{r_0} 
  \arctan \left( \frac{r}{r_0} \right) + \frac{Q^2}{4},
\label{sec3-100}\\
  \delta\Delta (r) &=& \frac{Q^2}{2} 
  \left[
  \frac{r}{r_0} \arctan \left( \frac{r}{r_0} \right) - 
  \frac{r^2 - r_0^2}{r^2 + r_0^2} + \frac{1}{2} 
  \right] .
\label{sec3-110}
\end{eqnarray} 
Substituting into equation \eqref{sec2-90} gives us 
\begin{equation}
  a(r) + \Delta (r) e^{2\psi(r)} = 2r_0^2 + \frac{Q^2}{2}
\label{sec3-120}
\end{equation}
but 
\begin{equation}
  a(r) \approx r_0^2 + r^2 - \frac{Q^2}{2} \frac{r}{r_0} 
  \arctan \left( \frac{r}{r_0} \right) + \frac{Q^2}{4}
\label{sec3-130}
\end{equation}
and 
\begin{equation}
  a(0) \approx r_0^2 + \frac{Q^2}{4}
\label{sec3-140}
\end{equation}
therefore 
\begin{equation}
  a(r) + \Delta (r) e^{2\psi(r)} = 2a(0)
\label{sec3-125}
\end{equation}
with an accuracy of $Q^2$. 

\section{Solutions expanded into a series}
\label{series}

In this section we will derive the solution of Eq's set \eqref{sec1-3}-\eqref{sec1-7} 
by expansion in terms of $r$. 
Using the MAPLE package we have obtained the solution with accuracy of 
$r^{12}$
\begin{eqnarray}
  \psi(x) &=& \frac{1}{4}{\tilde{Q}}^{2}{x}^{2} + 
  \left (
  {\frac {5}{48}} {\tilde{Q}}^{4} - \frac{1}{4} {\tilde{Q}}^{2}
  \right ){x}^{4} + 
  \left (
  -{\frac {7}{30}}{\tilde{Q}}^{4} + {\frac {41}{720}}{\tilde{Q}}^{6} + 
  \frac{1}{4} {\tilde{Q}}^{2}
  \right ){x}^{6}
\nonumber\\
  && + \left (-\frac{1}{4}{\tilde{Q}}^{2}+{\frac {213}{560}} {\tilde{Q}}^{4}-{
  \frac {111}{560}} {\tilde{Q}}^{6}+{\frac {281}{8064}} {\tilde{Q}}^{8}
  \right ){x}^{8}+
\nonumber\\
  && \left (-{\frac {341}{630}} {\tilde{Q}}^{4} + 
  {\frac {3749}{8400}} {\tilde{Q}}^{6} - {\frac {1553}{9450}} {\tilde{Q}}^{8}+
  {\frac {5147}{226800}} {\tilde{Q}}^{10} + \frac{1}{4}{\tilde{Q}}^{2}\right ){x}^
  {10}+O\left ({x}^{12}\right ),
\label{sec4-10}\\
  \Delta(x) &=& 
  1-{x}^{2}+
  \left (\frac{5}{6} {\tilde{Q}}^{2}-\frac{1}{4} {\tilde{Q}}^{4}
  \right ){x}^{4}+
  \left (
  -{\frac {9}{10}} {\tilde{Q}}^{2}+{\frac {47}{90}} {\tilde{Q}}^{4}-\frac{1}{12} {\tilde{Q}}^{6}
  \right ){x}^{6} + 
\nonumber\\ 
  && \left (-\frac{6}{7} {\tilde{Q}}^{4}+
  {\frac {373}{1260}} {\tilde{Q}}^{6}-{\frac {3}{80}} {\tilde{Q}}^{8}+{
  \frac {13}{14}} {\tilde{Q}}^{2}\right ){x}^{8} + 
\nonumber\\ 
  && \left ({\frac {218}{175}} {\tilde{Q}}^{4}-
  {\frac {2893}{4200}} {\tilde{Q}}^{6}+{\frac {5219}{
  28350}} {\tilde{Q}}^{8}-{\frac {17}{18}} {\tilde{Q}}^{2}-{\frac {37}{1890}} {\tilde{Q}}^{10}
  \right ){x}^{10} + O\left ({x}^{12}\right ),
\label{sec4-30}\\ 
  a(x) &=& 
  1+\left (1-\frac{1}{2} {\tilde{Q}}^{2}\right )
  \left[
  {x}^{2} + \frac{1}{6} {\tilde{Q}}^{2}{x}^{4} + 
  \left (-\frac{1}{10} {\tilde{Q}}^{2}+{\frac {11}{180}} {\tilde{Q}}^{4}\right ){x}^{6} + 
  \right .
\nonumber \\
  &&\left .
  \left (-{\frac {13}{140}} {\tilde{Q}}^{4}+{\frac {73}
  {2520}} {\tilde{Q}}^{6}+\frac{1}{14} {\tilde{Q}}^{2}\right ){x}^{8} + 
  \right .
\nonumber \\
  &&\left .  
  \left ({\frac {239}{2100}} {\tilde{Q}}^{4}+{\frac {887}{56700}} 
  {\tilde{Q}}^{8}-\frac{1}{18} {\tilde{Q}}^{2}-{\frac {13}{175}} {\tilde{Q}}^{6}
  \right ){x}^{10}
  \right] + O\left ({x}^{12}\right ) 
\label{sec4-40}
\end{eqnarray}  
here we have introduced the following dimensionless parameter : 
$\tilde{Q} = Q/\sqrt{a(0)}$. The substitution in Eq. \eqref{sec2-90} shows that 
it is valid. 
\par 
One of the most important question here is : how long is the $\Delta-$string ? 
Let us determine the length of the $\Delta-$string as $2 r_H$ where $r_H$ is 
the place where $ds^2 (\pm r_H) = 0$ (at these points $\Delta (\pm r_H) = 0$). 
This is very complicated problem because we have 
not the analytical solution and the numerical calculations indicate that 
$r_H$ depends very weakly from $\delta = (1 - Q/q)$ parameter : 
the big magnitude of this parameter leads to the relative small magnitude of 
$r_H$. The expansion \eqref{sec4-40} allows us to assume that 
\begin{equation}
  a(r) = a(0) +  a_1(r, Q) \, \delta_1
\label{sec4-50}
\end{equation}
here $\delta_1 = 1 - Q^2/Q_0^2$; $Q_0^2 = 2 a(0)$. If 
$Q \approx Q_0 = \sqrt{2a(0)}$ then 
\begin{equation}
  a(r) \approx a(0) +  a_1(r, Q_0) \, \delta_1 .
\label{sec4-60}
\end{equation}
Using the MAPLE package we have obtained $a_1(r)$ with accuracy of $r^{20}$ 
\begin{equation}
\begin{split}
  a_1(x) = {x}^{2} + \frac{1}{3} {x}^{4} + 
  {\frac {2}{45}} {x}^{6} + 
  {\frac {1}{315}} {x}^{8} + 
  {\frac {2}{14175}} {x}^{10} +  
\\
  {\frac {2}{467775}} {x}^{12} + 
  {\frac {4}{42567525}} {x}^{14} + 
  {\frac {1}{638512875}} {x}^{16} +
  {\frac {2}{97692469875}} {x}^{18} + 
  O\left ({x}^{20}\right ) . 
\label{sec4-70}
\end{split}
\end{equation}
It is easy to see that this series coincides with $\cosh^2(x)-1$. 
Therefore we can assume that in the first rough approximation 
\begin{equation}
  a(r) \approx a(0) 
  \left \{ 1 + 
  \left(
  1 - \frac{Q^2}{Q_0^2}
  \right)
  \left[
  \cosh^2 \left( \frac{r}{\sqrt{a(0)}}\right) - 1 
  \right] 
  \right \}.
\label{sec4-80}
\end{equation}
It allows us to estimate the length $L = 2 r_H$ of the $\Delta-$string from 
equation \eqref{sec2-90} as 
\begin{equation}
  a(r_H) = a(0) 
  \left \{
  1 + \delta_1 \left[ \cosh^2 \left( \frac{r_H}{\sqrt{a(0)}}\right)  - 1 \right]
  \right \} = 2a(0) .
\label{sec4-90}
\end{equation}
For $r_H \gg \sqrt{a(0)}$ we have 
\begin{equation}
  L \approx \sqrt{a(0)} \ln \frac{1}{\delta_1} .
\label{sec4-100}
\end{equation}
One can compare this result with the numerical calculations : 
$\delta_1 = 10^{-10}$, $r_H/\sqrt{a(0)} \approx 15$, 
$\frac{1}{2}\ln 1/\delta_1 \approx 15$; 
$\delta_1 = 10^{-9}$, $r_H/\sqrt{a(0)} \approx 11$, 
$\frac{1}{2}\ln 1/\delta_1 \approx 10$; 
$\delta_1 = 10^{-8}$, $r_H/\sqrt{a(0)} \approx 10$, 
$\frac{1}{2}\ln 1/\delta_1 \approx 9$. 
We see that $r_H/\sqrt{a(0)}$ and $\ln 1/\delta_1$ have the same order. 
This evaluation should be checked up (in future investigations) more 
carefully as the convergence radius of our expansion 
\eqref{sec4-10}-\eqref{sec4-40} is unknown. 

\section{The qualitative model of the $\Delta-$string}
\label{qualitative}

\begin{figure}[h]
  \begin{center}
  \fbox{
  \includegraphics[height=5cm,width=7cm]{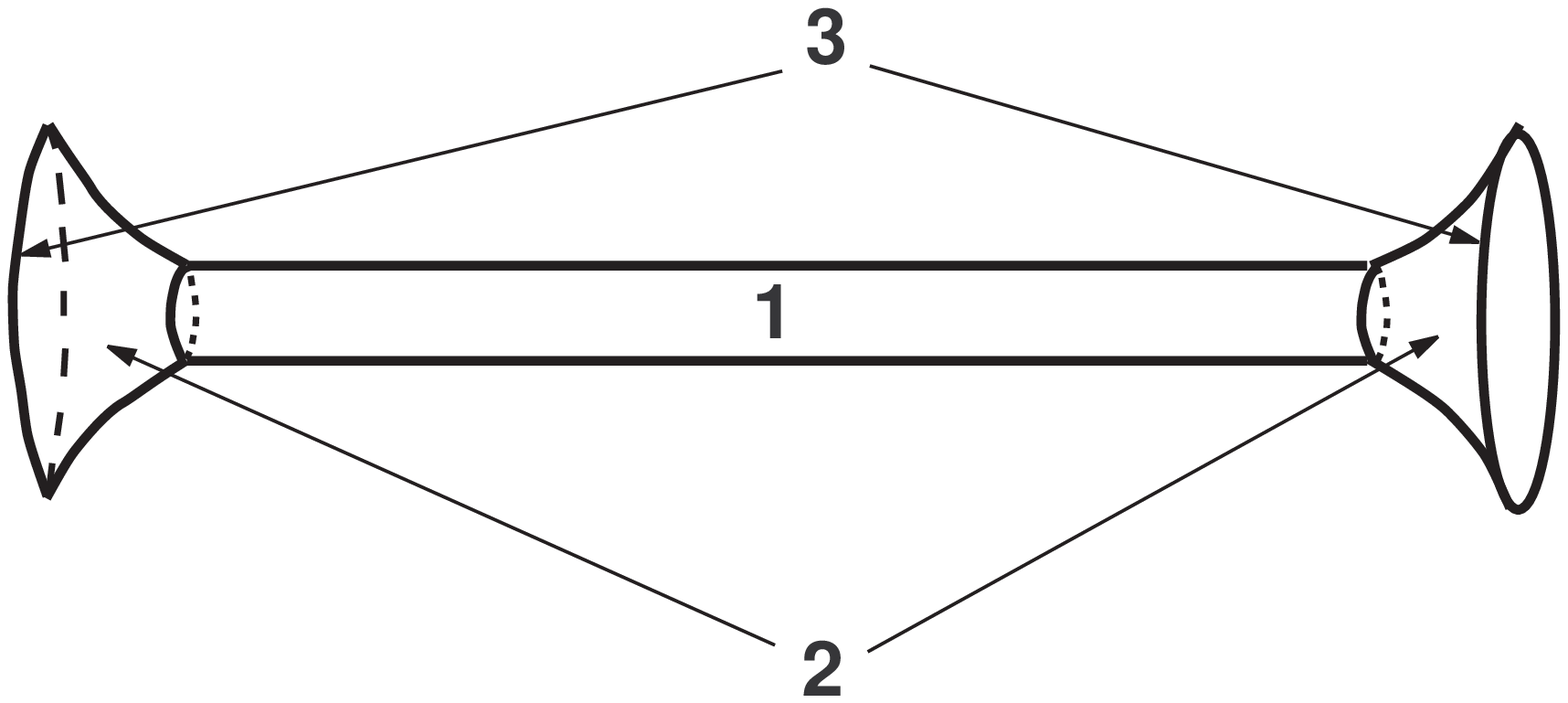}}
  \caption{The qualitative model of the $\Delta-$string. 
  \textbf{1} is the part of the infinite flux tube solution. 
  \textbf{2} are two cones which are pure electric solutions, 
  \textbf{3} are two hypersurfaces where $ds^2=0$.}
  \label{fig5}
  \end{center}
\end{figure}

The numerical calculations presented on Fig's \eqref{fig2}-\eqref{fig4} 
show that the $\Delta-$string approximately can be presented as a finite tube 
with the constant cross section and a big length joint at the ends with two short 
tubes which have variable cross section, see Fig. \eqref{fig5}. 
\par 
For the model of the central tube we take the part of the infinite flux tube 
solution with $Q = q = Q_0$ \cite{dzhsin1} 
\begin{eqnarray} 
a_\infty = \frac{Q_0^2}{2} = const, 
\label{sec5-10}\\
e^{\psi} = \frac{a_\infty}{\Delta} = \cosh\frac{r}{\sqrt{a_\infty}},
\label{sec5-20}\\
\omega = \sqrt{2}\sinh\frac{r}{\sqrt{a_\infty}} 
\label{sec5-30}
\end{eqnarray} 
here we have parallel electric $E$ and magnetic $H$ fields with equal electric 
$q$ and magnetic $Q$ charges. 
\par 
For a model of two peripheral ends (cones) we take the solution with $Q=0$ 
(equations \eqref{sec3-40}-\eqref{sec3-60}). Here we have only the electric 
field $E$. At the ends of the $\Delta-$string ($r = \pm r_H$) 
the function 
$\Delta(\pm r_H) = 0$ and the term $Q^2\Delta e^{2\psi}/a^3$ in equations 
\eqref{sec1-3}-\eqref{sec1-7} is zero that allows us to set this term as zero 
at the peripheral cones. 
\par 
We have to join the components of metric on the $r=r_1=r_H-\sqrt{a(0)}$ 
(here $a(0) = a_\infty$). 
\begin{equation}
  \cosh^2 \left( \frac{r_1}{\sqrt{a(0)}} \right) = 
  \frac{\tilde{a}(r_1)}{\tilde{\Delta}(r_1)} = 
  \tilde{g}_{tt}(r_1) = 
  \tilde{\tilde{g}}_{tt}(r_1) = 
  \frac{\tilde{\tilde{a}}(r_1)}{\tilde{\tilde{\Delta}}(r_1)}
\label{sec5-40}
\end{equation}
here $\tilde{ }$ and $\tilde{\tilde{ }}$ mean that the corresponding quantities 
are the metric components for $Q=Q_0$ and $Q=0$ solutions respectively. According 
to equations \eqref{sec3-40}-\eqref{sec3-60} 
\begin{eqnarray}
  \tilde{\tilde{a}}(r) &=& a(0) + \left( r - r_1 \right)^2 , 
\label{sec5-50}\\
  \tilde{\tilde{\Delta}}(r) &=& \alpha 
  \left[
  a(0) - \left( r - r_1 \right)^2 
  \right]
\label{sec5-60}\\ 
  \tilde{\tilde{a}}(r_1) &=& \tilde{a}(r_1) = a(0) 
\label{sec5-65}
\end{eqnarray}  
here $r_0^2$ from equation \eqref{sec3-40}-\eqref{sec3-60} is replaced with 
$\tilde{\tilde{a}}(0)$ and some coefficient $\alpha$ is introduced as Eq's 
\eqref{sec1-3}-\eqref{sec1-7} with $Q=0$ have the terms like 
$\Delta'' /\Delta$ and $\Delta'/\Delta$ only. It gives us 
\begin{equation}
  \alpha = \frac{1}{\cosh^2\left( \frac{r_1}{\sqrt{a(0)}} \right)} 
  \approx 4e^{\frac{-2 r_1}{\sqrt{a(0)}}}
\label{sec5-70}
\end{equation}
since for the long $\Delta-$string $r_1/\sqrt{a(0)} \gg 1$. The next joining 
is for 
\begin{equation}
  1 = \frac{\tilde{\Delta}(r_1) e^{2 \tilde{\psi}(r_1)}}{\tilde{a}(r_1)} = 
  \tilde{g}_{55}(r_1) = \tilde{\tilde{g}}_{55}(r_1) = 
  \frac{\tilde{\tilde{\Delta}}(r_1) e^{2 \tilde{\tilde{\psi}}(r_1)}}
  {\tilde{\tilde{a}}(r_1)}
\label{sec5-80}
\end{equation}
here the constant term $2 \tilde{\tilde{\psi}}_1$ is added to solution 
\eqref{sec3-60} $2\psi_0=0$ since again equations \eqref{sec1-3}-\eqref{sec1-7} 
have $\psi''$ and $\psi'$ terms only. Consequently 
\begin{equation}
  e^{2\tilde{\tilde{\psi}}_1} = \frac{1}{\alpha} \approx 
  \frac{e^{2 r_1/\sqrt{a(0)}}}{4} \quad 
  \text{or} \quad 
  \psi_1 \approx \frac{r_1}{\sqrt{a(0)}} - \ln 4 .
\label{sec5-90}
\end{equation}
The last component of the metric is 
\begin{equation}
  a(0) = a_\infty = \tilde{g}_{\theta\theta}(r_1) = 
  \tilde{\tilde{g}}_{\theta\theta}(r_1) = \tilde{\tilde{a}}(r_1) .
\label{sec5-100}
\end{equation}
For the electric field we have 
\begin{eqnarray}
  \tilde{E}(r_1) &=& \frac{q e^{\tilde{\psi}(r_1)}}{a(0)} = 
  \frac{q}{a(0)} \cosh \left( \frac{r_1}{\sqrt{a(0)}} \right) = 
  \frac{q}{a(0)} \frac{1}{\sqrt{\alpha}}
\label{sec5-110}\\
  \tilde{\tilde{E}}(r_1) &=& q \frac{\tilde{\tilde{a}}(r_1)}
  {\tilde{\tilde{\Delta}}^2(r_1)} 
  e^{-3 \tilde{\tilde{\psi}}(r_1)} = 
  \frac{q}{a(0)} \frac{1}{\sqrt{\alpha}}
\label{sec5-120}
\end{eqnarray}  
and consequently 
\begin{equation}
  \tilde{E}(r_1) = \tilde{\tilde{E}}(r_1) .
\label{sec5-130}
\end{equation}
Thus our final result for this section is that at the first rough approximation 
the $\Delta-$string looks like to a tube with constant cross section and two 
cones attached to its ends. 
\par 
In this section we have considered an approximate model of the $\Delta-$string. 
The purpose of this model is to show visually how is the form of the $\Delta-$string 
since it is not clear from the approximate calculations. From this point of view 
we do not demand the continuity of the metric derivatives as this model helps us 
only to understand what is it the $\Delta-$string on a qualitative level. 
Nevertheless we should not forget that the cross section of the long tube and 
all sizes of two peripheral cones are in Planck region and consequently quantum 
fluctuations can make unessential requirements to a continuity of classical 
degrees of freedom. 

\section{$\Delta-$string as a model of electric charge}
\label{charge}

In this section we would like to present a model of attachment of the 
$\Delta-$string to a spacetime. 
\par 
May be the most important question for the $\Delta-$string is : what will see 
an external observer living in the Universe to which the $\Delta-$string 
is attached. Whether he will see a dyon with electric and magnetic fields 
or an electric charge ? This question is not very simple as it is not very 
clear what is it the electric and magnetic fields on the $\Delta-$string. 
Are they the tensor $F_{\mu\nu}$ and fields $E_i$ and $H_i$ 
\begin{eqnarray}
  F_{\mu\nu} &=& \partial_\mu A_\nu - \partial_\nu A_\mu ,
  \quad \mu , \nu = 0,1,2,3 ,
\label{sec6-10}\\
  E_i &=& F_{0i}, \quad i,j,k = 1,2,3 ,
\label{sec6-20}\\
  H_i &=& e_{ijk} F^{jk} 
\label{sec6-30}  
\end{eqnarray}
or something another that is like to an electric displacement 
$D_i = \varepsilon E_i$ and $H_i = \mu B_i$ ? The answer on this question 
depends on the way how we will continue the solution behind the hypersurface 
$r=\pm r_H$ (where $ds^2=0$). We have two possibilities : the first 
way is the simple continuation our solution to $|r|>r_H$, the second way is a 
string approach - attachment the $\Delta-$string to a spacetime. In the 
first case $\Delta(r) > 0$ by $|r|<r_H$ and $\Delta(r) < 0$ by $|r|>r_H$ 
it means that by $|r|>r_H$ the time and 5th coordinate $\chi$ becomes 
respectively space-like and time-like dimensions. It is not so good for us. 
The second approach is like to a string attached to a D-brane. We believe that 
physically the second way is more interesting. 
\par 
Let us consider Maxwell equations in the 5D Kaluza-Klein theory 
\begin{equation}
  R_{5\mu} = \frac{1 }{\sqrt{-g}} \partial_\nu 
  \left(
  \sqrt{-g} g_{55} F^{\mu\nu}
  \right) = 0
\label{sec6-40}
\end{equation}
here $g = \det{g_{AB} = a^4 \sin^2 \theta e^{2\psi}}$. These 5D Maxwell 
equations are similar to Maxwell equations in the continuous medium where 
the factor $\sqrt{-g} g_{55}$ is like to a permittivity for 
$R_{50}$ equation and a permeability for $R_{5i}$ equations. It 
allows us offer the following conditions for matching the electric and magnetic 
fields at the attachment point of $\Delta-$ string to the spacetime 
\begin{eqnarray}
  \sqrt{-g_\Delta} {g_{\Delta}}_{55} F^{0i}_\Delta &=& \sqrt{-g} F^{0i} 
\label{sec6-50}\\
  \sqrt{-g_\Delta} {g_{\Delta}}_{55} F^{ij}_\Delta &=& \sqrt{-g} F^{ij} 
\label{sec6-60}
\end{eqnarray} 
here the subscript $\Delta$ means that this quantity is given on the 
$\Delta-$string and 
on l.h.s there are the quantities belonging to the $\Delta-$string 
and on the r.h.s. the corresponding quantities belong to the spacetime 
to which we want to attach the $\Delta-$string. As an example one can 
join the $\Delta-$string to the Reissner-Nordstr\"om solution with the 
metric 
\begin{equation}
  ds^2 = e^{2\nu} dt^2 - e^{-2\nu} dr^2 - r^2 
  \left( d\theta^2 + \sin^2\theta d\varphi^2 \right).
\label{sec6-70}
\end{equation}
Then on the $\Delta-$string 
\begin{equation}
  \sqrt{-g_\Delta} {g_{\Delta}}_{55} F^{tr}_\Delta = 
  - \frac{\Delta^2 e^{3\psi}}{a} \omega ' = 
  -a \frac{q}{a}
\label{sec6-80}
\end{equation}
For the Reissner-Nordstr\"om solution 
\begin{equation}
  \sqrt{-g} F^{tr} = - r^2 E_{r(RN)}.
\label{sec6-90}
\end{equation}
Therefore after joining on the event horizon $r=r_H$ ($a=a_H$) we have 
\begin{equation}
  E_{r(RN)} = \frac{q}{a_H}
\label{sec6-100}
\end{equation}
here we took into account that $a_H=r^2_H$. 
\par 
For equation \eqref{sec6-60} we have 
\begin{equation}
  \sqrt{-g_\Delta} {g_{\Delta}}_{55} F^{\theta \varphi}_\Delta = 
  \frac{Q \Delta e^{3\psi}}{a^2}.
\label{sec6-110}
\end{equation}
On the surface of joining $\Delta(r_H) = 0$ consequently we have very 
unexpected result : the continuation of $F^{\theta\varphi}$ 
(and magnetic field $H_r$) from the $\Delta-$string to the spacetime 
(D-brane) gives us zero magnetic field $H_r$. 
\par 
Finally our result for this section is that the $\Delta-$string can be 
attached to a spacetime by such a way that it looks like to an electric 
charge but not magnetic one. Such approach to the geometrical interpretation 
of electric charges suddenly explain why we do not observe magnetic charges 
in the world. 

\section{Waves on the $\Delta-$string}
\label{waves}

\par 
In this section we will consider small perturbations of 5D metric. Generally 
speaking they are 5D gravitational waves on the gravitational flux tube (or on 
the string language - vibrations of $\Delta-$string). In the general case 
5D metric is 
\begin{equation}
  ds^2 = g_{\mu\nu} dx^\mu dx^\nu - \phi^2
  \left(
  d\chi + A_\mu dx^\mu
  \right)^2
\label{sec1-56}
\end{equation}
here $g_{\mu\nu}$ is the 4D metric; $\mu , \nu = 0,1,2,3$; $A_\mu$ is the 
electromagnetic potential; $\phi$ is the scalar field. The corresponding 5D 
Kaluza-Klein's equations are  (for the reference see, for example, \cite{wesson}) 
\begin{eqnarray}
R_{\mu \nu} - \frac{1}{2} g_{\mu \nu} R & = & 
-\frac{\phi^2}{2} T_{\mu \nu} - \frac{1}{\phi} 
\left [
\nabla _\mu \left (
            \partial _\nu \phi
            \right ) - g_{\mu \nu} \Box \phi 
\right ] ,
\label{sec1-60}\\
\nabla _\nu F^{\mu \nu} & = & -3 \frac{\partial _\nu \phi}{\phi} F^{\mu \nu} ,
\label{sec1-70}\\
\Box \phi & = & - \frac{\phi^3}{4} F^{\alpha \beta} F_{\alpha \beta} 
\label{sec1-80}
\end{eqnarray}
where $R_{\mu \nu}$ is the 4D Ricci tensor; 
$F_{\mu \nu} = \partial_\mu A_\nu - \partial_\nu A_\mu$ is the 
4D Maxwell tensor and $T_{\mu \nu}$ is the energy-momentum tensor 
for the electromagnetic field. 
\par
We will consider only $\delta A_\mu$ perturbations, $\delta g_{\mu\nu}$ 
and $\delta \phi$ degrees of freedom are frozen. For this approximation 
we have equation 
\begin{equation}
  \nabla_\nu \delta F^{\mu\nu} = 0 .
\label{sec1-90}
\end{equation}
We introduce only one small perturbation in the electromagnetic potential 
\begin{equation}
  \delta A_\theta = f(t,r) 
\label{sec1-101}
\end{equation}
since for the background metric $\phi = const$. 
Generally speaking $A_\theta$-component should  have some dependence on 
the $\theta$-angle. But the cross section of the $\Delta-$string is in Planck region 
and consequently the points with different $\theta$ and $\varphi$ 
($r$ = const) physically are not distinguishable. Therefore all physical 
quantities on the $\Delta-$string should be averaged over polar angles 
$\theta$ and $\varphi$. It means that the $\delta A_\theta$ in 
Eq. \eqref{sec1-100} is averaged quantity. 
\par 
After such) remark we have the following wave equation 
for the function $f(t,r)$ 
\begin{equation}
  \partial_{tt} f(t,x) - \cosh x \partial_x 
  \biggl (
  \cosh x \partial_x f(t,x)
  \biggl ) = 0
\label{sec1-110}
\end{equation}
here we have introduced the dimensionless variables $t/\sqrt{a(0)} \rightarrow t$ 
and $r/\sqrt{a(0)} \rightarrow x$. The solution is 
\begin{equation}
  f(t,x) = f_0 F\left(t - 2 \arctan e^x \right) + 
  f_1 F\left(t + 2 \arctan e^x \right)
\label{sec1-121}
\end{equation}
here $f_{0,1}$ are some constants and $F$ and $H$ are arbitrary functions. 
This solution has more suitable form if we introduce new coordinate 
$y = 2 \arctan e^x$. Then 
\begin{equation}
  f(t,y) = f_0 F(t-y) + f_1 H(t + y).
\label{sec1-131}
\end{equation}
The metric is 
\begin{equation}
  ds^2 = \frac{a_0}{\sin^2y} dt^2 - \frac{dy^2}{\sin^2y} - 
  a_0\left( d\theta^2 + \sin^2\theta d\varphi^2 \right) - 
  \left(
  d\chi + \omega dt + Q \cos \theta d\varphi
  \right)^2 .
\label{sec1-140}
\end{equation}
Thus the simplest solution is electromagnetic waves moving in both 
directions along the $\Delta-$string. 

\section{The comparison with string theory}
\label{comparison}

Now we want to compare this situation with the situation in string theory. 
How is the difference between the result presented in Section \ref{waves} 
(Eq's \eqref{sec1-110}, \eqref{sec1-121}) 
and the string oscillation in the ordinary string theory ? The action for 
bosonic string is 
\begin{equation}
  S = - \frac{T}{2} \int d^2 \sigma \sqrt{h} h^{ab} 
  \partial_a \mathcal X^\mu \partial_b \mathcal X_\mu 
\label{sec2-10}
\end{equation}
here $\sigma^a = \{ \sigma , \tau \}$ is the coordinates on the world sheet of 
string; $\mathcal X^\mu$ are the string coordinates in the ambient spacetime; 
$h_{ab}$ is the metric on the world sheet. The variation with 
respect to $\mathcal X^\mu$ give us the usual 2D wave equation 
\begin{equation}
  \Box \mathcal X^\mu \equiv 
  \left(
  \frac{\partial^2}{\partial \sigma ^2} - 
  \frac{\partial^2}{\partial \tau ^2} 
  \right) \mathcal X^\mu = 0 
\label{sec2-20}
\end{equation}
which is similar to Eq. \eqref{sec1-110}. The difference is that the variation 
of the action \eqref{sec2-10} with respect to the metric $h^{ab}$ gives us 
some constraints equations in string theory 
\begin{equation}
  T_{ab} = \partial_a \mathcal X^\mu \partial_b \mathcal X_\mu - 
  \frac{1}{2} h_{ab} h^{cd} 
  \partial_c \mathcal X^\mu \partial_d \mathcal X_\mu = 0 
\label{sec2-30}
\end{equation}
but for $\Delta-$string analogous variation gives the dynamical equation 
for 2D metric $h_{ab}$. 
\par 
The more detailed description is the following. The topology of 5D 
Kaluza-Klein spacetime is $M^2 \times S^2 \times S^1$ 
where $M^2$ is the 2D space-time spanned on the time and longitudinal 
coordinate $r$; $S^2$ is the cross section of the flux tube solution and it is 
spanned on the ordinary spherical coordinates $\theta$ and $\varphi$; 
$S^1 = U(1)$ is the Abelian gauge group which in this consideration is 
the 5th dimension. The initial 5D action for 
$\Delta-$string is 
\begin{equation}
  S = \int \sqrt{-\stackrel{(5)}{g}} \stackrel{(5)}{R} 
\label{sec2-31}
\end{equation}
where $\stackrel{(5)}{g}$ is the 5D metric \eqref{sec1-56}; 
$\stackrel{(5)}{R}$ are 5D Ricci scalars. We want to reduce the initial 
5D Lagrangian \eqref{sec2-31} to a 2D Lagrangian. Our basic assumption is that 
the sizes of 5$^{th}$ and dimensions 
spanned on polar angles $\theta$ and $\varphi$ is approximately 
$\approx l_{Pl}$.  At first we have usual 5D $\rightarrow$ 4D 
Kaluza-Klein dimensional reduction. Following, for example, to review 
\cite{wesson} we have  
\begin{equation}
\frac{1}{16 \pi \stackrel{(5)}{G}}  \stackrel{(5)}{R} = 
\frac{1}{16 \pi G} \stackrel{(4)}{R} - \frac{1}{4} \phi^2 
F_{\mu \nu} F^{\mu \nu} + \frac{2}{3} 
\frac{\partial_\alpha \phi \partial^\alpha \phi}{\phi^2}
\label{sec2-40}
\end{equation}
where $\stackrel{(5)}{G} = G \int dx^5$ is 5D gravitational constant; 
$G$ is 4D gravitational constant; $\stackrel{(4)}{R}$ is the 4D Ricci scalars. 
The determinants of 5D 
and 4D metrics are connected as $\stackrel{(5)}{g} = \stackrel{(4)}{g}\phi$. 
One of the basis paradigm of quantum gravity is that a minimal 
length in the Nature is the Planck scale. Physically it means that not any 
physical fields depend on the 5$^{th}$, $\theta$ and $\varphi$ coordinates. 
\par 
The next step is reduction from 4D to 2D. The 4D metric can be expressed as 
\begin{eqnarray}
d\stackrel{(4)}{s^2} & = & g_{\mu \nu} dx^\mu dx^\nu = 
g_{ab}(x^c) dx^a dx^b + 
\nonumber \\
&&\chi(x^c)
\left (
\omega ^{\bar i}_i dy^i + B^{\bar i}_a(x^c) dx^a
\right )
\left (
\omega ^{\bar j}_j dy^j + B_{\bar i a}(x^c) dx^a
\right ) \eta_{\bar i \bar j}
\label{sec2-60}
\end{eqnarray}
where $a,b, c = 0,1$; $x^a$ are the time and longitudinal coordinates; \\
$-\left (\omega^{\bar i}_i dy^i \right ) 
\left (\omega ^{\bar j}_j dy^j \right ) \eta_{\bar i \bar j} = dl^2$ 
is the metric 
on the 2D sphere $S^2$; $y^i$ are the coordinates on the 2D sphere $S^2$; 
all physical quantities $g_{ab}, \chi$ and $B_{\bar i a}$ can depend only 
on the physical coordinates $x^a$. Accordingly to Ref. \cite{coq} we have 
the following dimensional reduction to 2 dimensions 
\begin{eqnarray}
\stackrel{(4)}{R} & = & \stackrel{(2)}{R} + R(S^2) - 
\frac{1}{4} \phi^{\bar i}_{ab} \phi^{ab}_{\bar i} - 
\nonumber \\
&&\frac{1}{2} h^{ij}h^{kl}
\left (
D_a h_{ik} D^a h_{jl} + D_a h_{ij} D^a h^{kl}
\right ) - 
\nabla^a 
\left (
h^{ij} D_a h_{ij}
\right )
\label{sec2-70}
\end{eqnarray}
where $\stackrel{(2)}{R}$ is the Ricci scalar of 2D spacetime; 
$D_\mu$ and $\phi^{\bar i}_{ab}$ are, respectively, the covariant derivative 
and the curvature of the principal connection $B^{\bar i}_a$ and 
$R(S^2)$ is the Ricci scalar of the sphere 
$S^2 =  \mathrm{SU(2)/U(1)}$ with 
linear sizes $\approx l_{Pl}$; $h_{ij}$ is the metric on $\mathcal L$ 
\begin{eqnarray}
\mathrm{su}(2) & = & \mathrm{Lie (SU(2))} = \mathrm{u}(1) \oplus \mathcal L ,
\label{sec2-80}\\
\mathrm{u}(1) & = &\mathrm{Lie(U(1))}
\label{sec2-100}
\end{eqnarray}
here $\mathcal L$ is the orthogonal complement of the u(1) algebra in the su(2) 
algebra; the index $i \in \mathcal L$. The metric $h_{ij}$ is 
proportional to the scalar $\chi$ in Eq. \eqref{sec2-30}. 
\par 
The situation with the electromagnetic 
fields $A_\mu$ and $F_{\mu \nu}$ is 
\begin{eqnarray}
A_\mu & = & \left \{ A_a, A_i \right \} \; 
A_a \; \text{is the vector;} \; 
A_i \; \text{are 2 scalars}; 
\label{sec2-85}\\
F_{ab} & = & \partial_a A_b - \partial_b A_a \; 
\text{is the Maxwell tensor for} \; A_a;
\label{sec2-90}\\
F_{ai} & = & \partial_a A_i ,
\label{sec2-105}\\
F_{ij} & = & 0
\end{eqnarray}
here we took into account that $\partial_i = 0$ as \textit{the point 
cannot be colored}. 
\par 
Connecting all results we see that only the following 
\textit{physical fields} on the $\Delta-$string are possible : 
2D metric $g_{ab}$, gauge fields $B^{\bar i}_a$, vectors $A_a$, 
tensors $F_{ab}, F_{a\bar i}, \Phi^{\bar i}_{ab}$, scalars $\chi$ and 
$\phi$. 
\par
The 2D action is 
\begin{equation}
  S = \int d^2 x^a \phi \left( \det g_{ab} \right) 
  \left( \det {\omega^{\bar i}_i} \right) 
  \left[
  \frac{\stackrel{(2)}{R}}{16 \pi G} - \frac{1}{4} \phi^2 
    \left(
    F_{t \theta} F^{t \theta} + F_{r \theta} F^{r \theta} 
    \right) + \text{other terms} 
  \right] .
\label{sec2-110}
\end{equation}
The second term in the $[\ldots ]$ brackets give us the wave equation 
\eqref{sec1-110} and the most important is that the first term is not total 
derivative in contrast with the situation in ordinary string theory in the 
consequence of the factors $\phi$ and 
$\left( \det {\omega^{\bar i}_i}\right)$. Therefore the variation with respect 
to 2D metric $g_{ab}$ give us some dynamical equations contrary to string theory 
where this variation leads to the constraint equations \eqref{sec2-30}. 
\par 
This remark allows us to say that the $\Delta-$string do not have such peculiarities 
as critical dimensions (D=26 for bosonic string). The reason for this is very 
simple : the comprehending space for $\Delta-$string is so small that it coincides 
with $\Delta-$string. Thus we can suppose that the critical dimensions 
in string theory is connected with the fact that the string curves the 
external space but the back reaction of curved space on the metric of 
the string world sheet is not taken into account. 

\section{Fermionic degrees of freedom on $\Delta-$string}
\label{fermionic}

The above-mentioned comparison of bosonic degrees of freedom on $\Delta$ and ordinary 
strings shows the similarity of dynamical behaviour of these quantities. But for strings
in string theory very essential are fermionic degrees of freedom. Therefore the 
question about the presence of fermionic fields on the $\Delta-$string becomes 
very important. Of course one can insert these fields in 5D Kaluza-Klein theory 
by hand and then search $\Delta-$string solutions in this theory. But from authors 
point of view it is not so fine as all bosonic degrees of freedom have natural geometrical 
origin (5D metric) but the fermionic degrees of freedom are some external fields. 
Fortunatelly there is an idea how fermions can be naturally and geometrically 
incorporated into gravity or more exactly into quantum gravity. 
\par
In Ref. \cite{smolin} Smolin have offered an idea that the wormholes in 
spacetime foam can be described as fermions. One can apply this idea for our goals 
by such a manner that fermionic degrees of freedom describe spacetime foam on 
$\Delta-$string. By such approach every quantum wormhole in spacetime foam is like 
to electric dipole in a continous medium as such wormhole can entrap the electric force 
lines. Let us describe Smolin's idea more careful. 

\section{The grassmanian operator of a minimalist wormhole}
\label{grassman}

Let us introduce an operator $W(x,y)$ which describes creation/annihilation 
minimalist wormhole connected two points $x$ and $y$ \cite{smolin} \cite{dzh1}. 
Let the operator $W(x,y)$ 
has the following property 
\begin{equation}
  W^2(x,y) = 0.
\label{sec1-10}
\end{equation}
It means that the reiterated creation/annihilation minimalist wormhole is 
senseless. If the square of some quantity is zero this quantity can be expressed  
wia Grassman numbers. Basically the operator $W(x,y)$ is nonlocal 
one but here we will consider the simplest case when $W(x,y)$ can be factorized 
on two local operators 
\begin{equation}
  W(x,y) = \theta(x) \theta(y).
\label{sec1-20}
\end{equation}
The Grassman number $\theta(x)$ can be considered as \textit{a readiness} of a point $x$ to 
pasting together with a point $y$ (or conversely the separation these points 
in the minimalist wormhole). The nonlocal function $W(x,y)$ in some sense is similar 
to 2-point Green's function in quantum field theory. The Green's functions 
$G(x_1, \ldots, x_n)$ describe the correlation between quantum fields in the points 
$x_1, \ldots, x_n$
\begin{equation}
	G\left(x_1, \ldots, x_n \right) = 
	\left\langle Q \left|
	  \phi\left( x_1 \right) \ldots \phi\left( x_n \right)
	\right| Q \right\rangle
\end{equation}
where $\left\langle Q \left| \right.\right.$ is a quantum state. In similar manner 
$W(x,y)$ describes such correlation between two points $x$ and $y$ that these points will 
be connected by a wormhole. 
\par
\begin{figure}[h]
\begin{center}
\fbox{
\includegraphics[height=5cm,width=5cm]{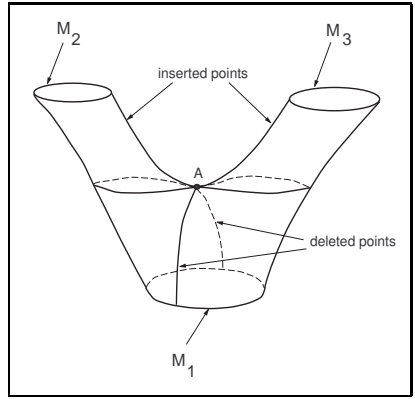}}
\caption{2 dimensional example of the topology change: $M_1$ is an initial 
space and $M_{2,3}$ are the final spaces which are obtained after some 
topological surgery. $A$ is the point where the time direction is not determined and the metric 
becomes singular.}
\label{mors}
\end{center}
\end{figure}

 
Very essential question appears in such situation: can be $W(x,y)$ operator 
deduced from some quantum theory of gravity by the quantization only field degrees of 
freedom (metric) ? This question is not trivial because appearing/disappearing of 
quantum wormholes occurs with the topology change. \textit{i.e.} every wormhole 
changes the topology of spacetime. Such change of topology takes place by deleting 
and inserting points into an initial space \cite{mors}, see Fig. \ref{mors}. 
But not any gravity theory working with field degrees of freedom can describe 
such process as on the deleted/inserted points the metric becomes singular. 
Thus the authors point of view is that \textit{the process of creation/annihilation of quantum 
handles (wormholes) have to be described by an independent manner which can not be 
deduced from the quantization of field degrees of freedom.}
\par
Let us consider three different presentation of $\theta(x)$ operator with scalar 
and spinor fields. 
\par 
\underline{The first case} is
\begin{equation}
  W^{ab}(x,y) = \phi(x) \theta^a \phi(y) \theta^b
\label{sec1-30}
\end{equation}
where $a,b$ are the undotted spinor indices, $\phi(x)$ is the scalar field, 
$\theta^a$ is the Grassman number 
\begin{equation}
  \theta^a \theta^b + \theta^b \theta^a = 0.
\label{sec1-40}
\end{equation}
It is easy to proof that 
\begin{equation}
\begin{split}
  W^{ab}(x,y) W_{ab}(x,y) = \phi^2(x) \phi^2(y) 
  \theta^a \theta^b \theta_a \theta_b & = 
\\  
  \phi^2(x) \phi^2(y) \varepsilon_{aa'}\varepsilon_{bb'}
  \theta^a \theta^b \theta^{a'}\theta^{b'} & \equiv 0 
\end{split}  
\label{sec1-55}
\end{equation}
where 
\begin{equation}
    \varepsilon^{ab} = 
    \begin{pmatrix}
    0 & 1 \\
    -1 & 0
  \end{pmatrix} ,
  \qquad 
  \varepsilon_{ab} = 
    \begin{pmatrix}
    0 & -1 \\
    1 & 0
  \end{pmatrix} .
\label{sec1-75}
\end{equation}
\par
{\underline{The second case} is
\begin{equation}
\begin{split}
  W(x,y) = \psi_a(x) \theta^a \psi_b(y) \theta^b = 
  \psi_a(x) \psi_b(y) \theta^a \theta^b & = 
\\  
  \left(
  \psi_1(x) \psi_2(y) - \psi_2(x) \psi_1(y)
  \right)
  \theta^1 \theta^2 &
\end{split}
\label{sec1-81}
\end{equation}
where $\psi_a(x)$ is an undotted spinor field in 
$(\frac{1}{2},0)$ representation. The square is zero
\begin{equation}
  W^2(x,y) = \left(
  \psi_1(x) \psi_2(y) - \psi_2(x) \psi_1(y)
  \right)^2
  \theta^1 \theta^2 \theta^1 \theta^2 \equiv 0.
\label{sec1-91}
\end{equation}
In this case $W(x,x) \equiv 0$. It means that the minimalist wormhole can 
connect only two different points. 
\par 
The same can be written for $(0, \frac{1}{2})$ representation. 
\par
\underline{The third case} is
\begin{equation}
  W(x,y) = \psi_a(x) \theta^a \psi_{\dot b}(y) \bar{\theta}^{\dot b} = 
  \psi_a(x) \psi_{\dot b}(y) \theta^a \bar{\theta}^{\dot b}
\label{sec1-100}
\end{equation}
where $\psi_{\dot b}$ is a dotted spinor in $(0,\frac{1}{2})$ representation. 
The square is zero 
\begin{equation}
\begin{split}
  W^2(x,y) = \left( 
  \psi_a(x) \psi_{\dot b}(y) \theta^a \bar{\theta}^{\dot b}
  \right)
  \left( 
  \psi_c(x) \psi_{\dot d}(y) \theta^c \bar{\theta}^{\dot d}
  \right) & = 
\\  
  \Bigl(
  \psi_a(x) \psi_c(x) \theta^a \theta^c 
  \Bigr)
  \left(
  \psi_{\dot b}(y) \psi_{\dot d}(y) \bar{\theta}^{\dot b} \bar{\theta}^{\dot d}
  \right) & \equiv 0
\end{split}  
\label{sec1-111}
\end{equation}
as 
\begin{equation}
  \psi_a(x) \psi_c(x) \theta^a \theta^c = 
  \psi_1(x) \psi_2(x) \theta^1 \theta^2 + 
  \psi_2(x) \psi_1(x) \theta^2 \theta^1 \equiv 0.
\label{sec1-120}
\end{equation}
The third case is more interesting because it has both dotted and undotted 
spinors. 
We have to note that as the classical canonical theory cannot describes topology 
change these operators do not correspond to any classical observables. 
It corresponds to the well-known fact that the Grassman numbers do not have 
any classical interpretation. 
\par 
We can suppose that $W(x,y)$ operator is described with dynamical fields : 
scalar field $\phi(x)$ or spinor field $\psi(x)$. The problem here is what kind of 
equations we have to apply for the definition of $\phi /\psi$ dynamic ? 
\par 
The operator $W(x,y)$ can be connected with an indefiniteness 
(the loss of information) of our knowledge about two points $x^\mu$ and 
$y^\mu$~: we do not know if these points are connected by the 
quantum minimalist wormhole or not. 
The physical sense of $W(x,y)$ operator is that 
\begin{equation}
  W(x,y) \approx 0 
  \quad \text{for} \quad 
  |x - y| >l_0 
\label{sec1-130}
\end{equation}
where $l_0$ is some characteristic length depending on some external conditions. 
For example, for the ordinary spacetime foam without any strong external 
fields $l_0 = l_{Pl} = 10^{-33}cm$. But in the presence of an external electric field 
$l_0$ can be changed as quantum wormholes can entrap the force lines of electric 
field. 
\par 
Let us introduce (following to Smolin) an infinitesimal operator 
$\delta B(x^\mu \rightarrow {x'}^\mu)$ of a displacement of the wormhole mouth 
\begin{eqnarray}
    \delta B 
    \biggl(
    \theta^{a} \rightarrow {(\theta^{a})}', 
    \bar{\theta}^{\dot{a}} \rightarrow {(\bar{\theta}^{\dot{a}})}',
		x^\mu & \rightarrow & {x'}^\mu 
		\biggl)
    W(y^\mu , x^\mu) = 
    W(y^\mu , {x'}^\mu ) ,
 \label{sec3.2-12a}\\
    W(y^\mu , x^\mu) & = & 
    \psi_a(y^\mu) \psi_{\dot b}(x^\mu) \theta^a \bar{\theta}^{\dot b} ,
 \label{sec3.2-12b}\\
    W(y^\mu , {x'}^\mu) & = & 
    \psi_a(y^\mu) \psi_{\dot b}({x'}^\mu) 
    \left( \theta^a + \varepsilon^a \right)
    \left( \bar{\theta}^{\dot b} + \bar{\varepsilon}^{\dot b} \right) 
 \label{sec3.2-12c}
\end{eqnarray}
here $\varepsilon^\alpha, \bar{\varepsilon}^{\dot b}$ is infinitesimal 
Grassmannian numbers. Therefore we have the following equation for the definition of 
$\delta B(z^A \rightarrow {z'}^A)$ 
operator (here $z^A=\{ x^\mu, \theta^a, \bar{\theta}^{\dot b} \}$ is the coordinates on 
a superspace)
\begin{equation}
    \delta B\left(z^A \rightarrow {z'}^A \right) 
    \psi_a(y^\mu) \psi_{\dot b}(x^\mu) \theta^a \bar{\theta}^{\dot b}= 
    \psi_a(y^\mu) \psi_{\dot b}({x'}^\mu) 
    \left( \theta^a + \varepsilon^a \right)
    \left( \bar{\theta}^{\dot b} + \bar{\varepsilon}^{\dot b} \right)  .
\label{sec3.2-12d}
\end{equation}
This equation has the following solution 
\begin{equation}
    \delta B\left(z^A \rightarrow {z'}^A \right) =  
    1 + \varepsilon^\alpha \frac{\partial}{\partial \theta^\alpha} - 
    i\varepsilon^\alpha \sigma^\mu_{\alpha\dot{\beta}} 
    \bar{\theta}^{\dot{\beta}} \partial_\mu + 
    \bar{\varepsilon}^{\dot\alpha} 
    \frac{\partial}{\partial \bar{\theta}^{\dot\alpha}} + 
    i\theta^\alpha \sigma^\mu_{\alpha\dot{\beta}} 
    \bar{\varepsilon}^{\dot{\beta}} \partial_\mu
\label{sec3.2-12e}
\end{equation}
For the proof of Eq. \eqref{sec3.2-12e} we shall calculate 
the effect of the $\delta B$ operator on the $W(y^\mu, x^\mu)$ operator. 
On the one hand we have 
\begin{equation}
\begin{split}
  &\delta B\left(z^A \rightarrow {z'}^A \right) 
  W \left( y^\mu , x^\mu \right) = \\
  &\biggl(1 + \varepsilon^\alpha \frac{\partial}{\partial \theta^\alpha} - 
    i\varepsilon^\alpha \sigma^\mu_{\alpha\dot{\beta}} 
    \bar{\theta}^{\dot{\beta}} \partial_\mu + 
    \bar{\varepsilon}^{\dot\alpha} 
    \frac{\partial}{\partial \bar{\theta}^{\dot\alpha}} + 
    i\theta^\alpha \sigma^\mu_{\alpha\dot{\beta}} 
    \bar{\varepsilon}^{\dot{\beta}} \partial_\mu
  \biggl)
  \psi_a(y^\mu) \psi_{\dot b}(x^\mu) \theta^a \bar{\theta}^{\dot b} = \\
  &\psi_a(y^\mu) \psi_{\dot b}(x^\mu) \theta^a \bar{\theta}^{\dot b} + 
  \psi_a(y^\mu) \psi_{\dot b}(x^\mu) \varepsilon^a \bar{\theta}^{\dot b} + 
  \psi_a(y^\mu) \psi_{\dot b}(x^\mu) \theta^a \bar{\varepsilon}^{\dot b} + \\
  &\psi_a(y^\mu) \psi_{\dot b}(x^\mu) \delta x^\mu \theta^a \bar{\theta}^{\dot b} 
  \approx 
  \psi_a(y^\mu) \psi_{\dot b}(x^\mu) 
  \left( \theta^a + \varepsilon^a \right)
  \left(\bar{\theta}^{\dot b} + \bar{\varepsilon}^{\dot b} \right) + \\
  &\psi_a(y^\mu) \psi_{\dot b}(x^\mu) \delta x^\mu \theta^a \bar{\theta}^{\dot b}
\label{app1}
\end{split}
\end{equation}
where 
\begin{equation}
  \delta x ^\mu = 
  -i\varepsilon^\alpha \sigma^\mu_{\alpha\dot{\beta}} 
  \bar{\theta}^{\dot{\beta}} + 
  i\theta^\alpha \sigma^\mu_{\alpha\dot{\beta}} 
  \bar{\varepsilon}^{\dot{\beta}} . 
\label{app2a}
\end{equation}
On the other hand 
\begin{equation}
\begin{split}
  &\delta B\left(z^A \rightarrow {z'}^A \right) 
  W \left( y^\mu , x^\mu \right) = 
  \psi_a(y^\mu) \psi_{\dot b}(x^\mu + \delta x^\mu) 
  \left( \theta^a \right)' 
  \left(\bar{\theta}^{\dot b}\right)' = \\
  &\psi_a(y^\mu) \psi_{\dot b}(x^\mu ) 
  \left( \theta^a \right)' 
  \left(\bar{\theta}^{\dot b}\right)' + 
  \psi_a(y^\mu) \psi_{\dot b}(x^\mu) \delta x^\mu \theta^a \bar{\theta}^{\dot b}
\label{app2}
\end{split}
\end{equation}
where $\left( \theta^a \right)' = \theta^a + \epsilon^a$ and 
$\left(\bar{\theta}^{\dot b}\right)' = \bar{\theta}^{\dot b} + \bar{\epsilon}^{\dot b}$. 
We see that r.h.s. of Eq's \eqref{app1} and \eqref{app2a} coincide. 
It means that the operator 
$W^{\gamma\delta}( y^\mu , {x'}^\mu )$ with the shifted 
wormhole mouth is equivalent to the shift of the coordinates in superspace 
$z^A \rightarrow {z'}^A$. 
\par 
After this we can say that 
$\theta = \{ \theta^\alpha , \theta^{\dot\alpha} \}$ 
are the Grassmanian numbers which we should use as some additional 
coordinates for the description of the spacetime foam. 
\par 
Such approach can give us an excellent possibility for understanding 
of geometrical meaning of spin-$\hbar /2$. Wheeler \cite{wheel} 
has mentioned repeatedly the importance of a geometrical interpretation 
of spin-$\hbar /2$. He wrote: \textit{the geometrical description 
of $\hbar /2$-spin must be a significant component of any electron model}. 

\section{Geometrical interpretation}
\label{geometry}

In this approach \textit{superspace is a model of the spacetime foam in quantum gravity}. 
The indefiniteness connected with the creation/annihilation of quantum
minimalist wormholes is described by Grassmannian coordinates (see,
Fig.\ref{fig5a}). In this interpretation an infinitesimal
Grassmannian coordinate transformation is associated with 
a displacement of the wormhole mouth, \textit{i.e.} with a 
change of the identification procedure (see, Fig.\ref{fig5a})
\begin{equation}
  \theta' = \theta + \varepsilon =
  Id\biggl( y^\mu \;
  \raisebox{1.5ex}{$\underleftrightarrow{
  \scriptstyle{\;\;yes \; or \; no \; ?\;\;}}$}
  \;
  {x'}^\mu = \\
  x^\mu - i\varepsilon^\alpha \sigma^\mu_{\alpha\dot{\beta}}
  \bar{\theta}^{\dot{\beta}} +
  i\theta^\alpha \sigma^\mu_{\alpha\dot{\beta}}
  \bar{\varepsilon}^{\dot{\beta}} 
  \biggl) = W\left( y^\mu , {x'}^\mu \right) 
\label{sec1-21}
\end{equation}
here the identification procedure $Id(\ldots)$ 
is described by the operator 
$W\left( y^\mu , {x'}^\mu \right)$. 
In this case the Grassmannian coordinate transformation
has a very clear geometrical sense : it describes a displacement 
of the wormhole mouth or it is the change of the
identification prescription. It is necessary to note that in 
Ref. \cite{gozzi} there is a similar interpretation of the 
Grassmanian ghosts : they are the Jacobi fields which are the 
infinitesimal displacement between two classical trajectories. 
\par
In this geometrical approach supersymmetry means that a
supersymmetrical Lagrangian is invariant under the identification
procedure \eqref{sec1-10}, \textit{i.e.}
\textit{the corresponding supersymmetrical
fields must be described in an invariant manner on the background of
the spacetime foam.}
\begin{figure}
\begin{center}
\fbox{
\includegraphics[height=5cm,width=5cm]{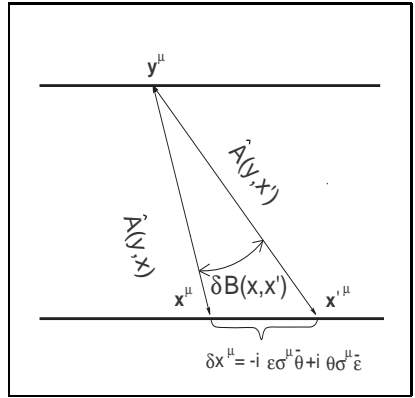}}
\caption{The distinction between two identification prescriptions
$\theta$ and $\theta'$ leads to a displacement $\delta x^\mu =
-i\varepsilon^\alpha \sigma^\mu_{\alpha\dot{\beta}}
  \bar{\theta}^{\dot{\beta}} +
  i\theta^\alpha \sigma^\mu_{\alpha\dot{\beta}}
  \bar{\varepsilon}^{\dot{\beta}}$}
\label{fig5a}
\end{center}
\end{figure}

\section{Fermionic degrees of freedom on the $\Delta-$string}

The fermionic degrees of freedom can appear on the $\Delta-$string if we suppose 
that quantum gravity play an essential role on the $\Delta-$string. It is the 
natural assumption because the cross section of the $\Delta-$string is in 
Planck region. It means that the spacetime foam should give rise to an essential 
contribution for the consideration of the $\Delta-$string characteristics. 
Taking into account the presence of quantum wormholes of spacetime foam one can 
imagine the $\Delta-$string as a shaggy $\Delta-$string, see Fig. \ref{shaggy}. 
\begin{figure}
\begin{center}
\fbox{
\includegraphics[height=5cm,width=5cm]{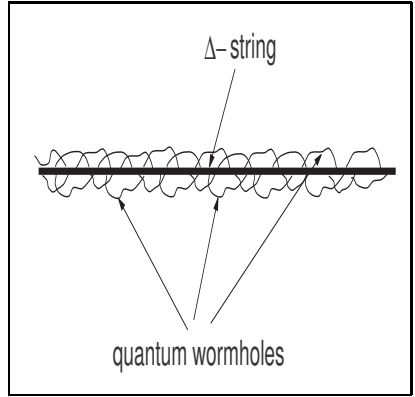}}
\caption{The shaggy $\Delta-$string dressed with quantum minimalist wormholes 
which can be described by a fermion field.}
\label{shaggy}
\end{center}
\end{figure}
\par 
Taking into account Smolin's idea about connection between fermions and topology 
(minimalist quantum wormholes in spacetime foam) one can suppose that spacetime 
foam on the $\Delta-$string generates fermionic degrees of freedom and the reasonongs 
similar to Section \ref{grassman} will lead to an object which is similar to 
superstrings in string theory. 
\par 
The problem here is that we should determine what kind of the dynamic (Lagrangian) 
we should use for spinor field $\psi$ in Eq's \eqref{sec1-80} \eqref{sec1-100} ?
According to Section \ref{geometry} it can be supergravity where we have gravitation, 
gauge fields and their fermionic partners. In this approach one can investigate 
gravitational flux tube solutions with the presence of fermionic fields. Unfortunatelly 
in supergravity the solutions with spinor fields almost are unknown. Evidently it is 
connected with mathematical problems of the corresponding Rarita-Schwinger equation. 
Thus it can be the problem for the future investigations. 

\section{Conclusions}

In this letter we have considered the gravitational flux tube solutions in 5D 
Kaluza-Klein gravity. These solutions are labeled by the relation between 
electric and magnetic fields. We have shown that in the case $E \approx H$ 
(with $E > H$) the corresponding spacetime has a part which is like to a flux 
tube filled with the electric and magnetic fields. Such tube can be arbitrary 
thin and long. It allows us to call such superthin and superlong tubes as a string-like 
objects, namely a $\Delta-$string. We have shown that taking into account 
the infinitesimal of the cross section (which is in the Planck region) one can 
reduce 5D initial degrees of freedom to 2D degrees of freedom living on the 
$\Delta-$string. The corresponding equations describe the propagation of scalar 
and gauge fields through the $\Delta-$string. It is necessary to note that 
2D dynamic on the $\Delta-$string differs from the dynamic of the ordinary string in 
string theory. The reason is very simple: $\Delta-$string has non-zero cross 
section. 
\par 
On the $\Delta-$string one can introduce fermionic degrees of freedom if we take into 
account such phenomena in quantum gravity as spacetime foam. According to Smolin 
the quantum wormholes can be connected with a fermion field which describes 
minimalist wormholes as it presented in Section \ref{grassman}, \ref{geometry}. 
In this model every quantum wormhole of spacetime is a short $\Delta-$string 
attached to an external spacetime, see Fig. \ref{qwormholes}. In such scenario 
quantum wormholes (or Smolin's minimalist wormholes) are described by a fermion field. 
According to Sections \ref{grassman}, \ref{geometry} these fermion fields are 
fermion ingredients of some supergravity theory. 
\begin{figure}[h]
\begin{center}
\fbox{
\includegraphics[height=5cm,width=5cm]{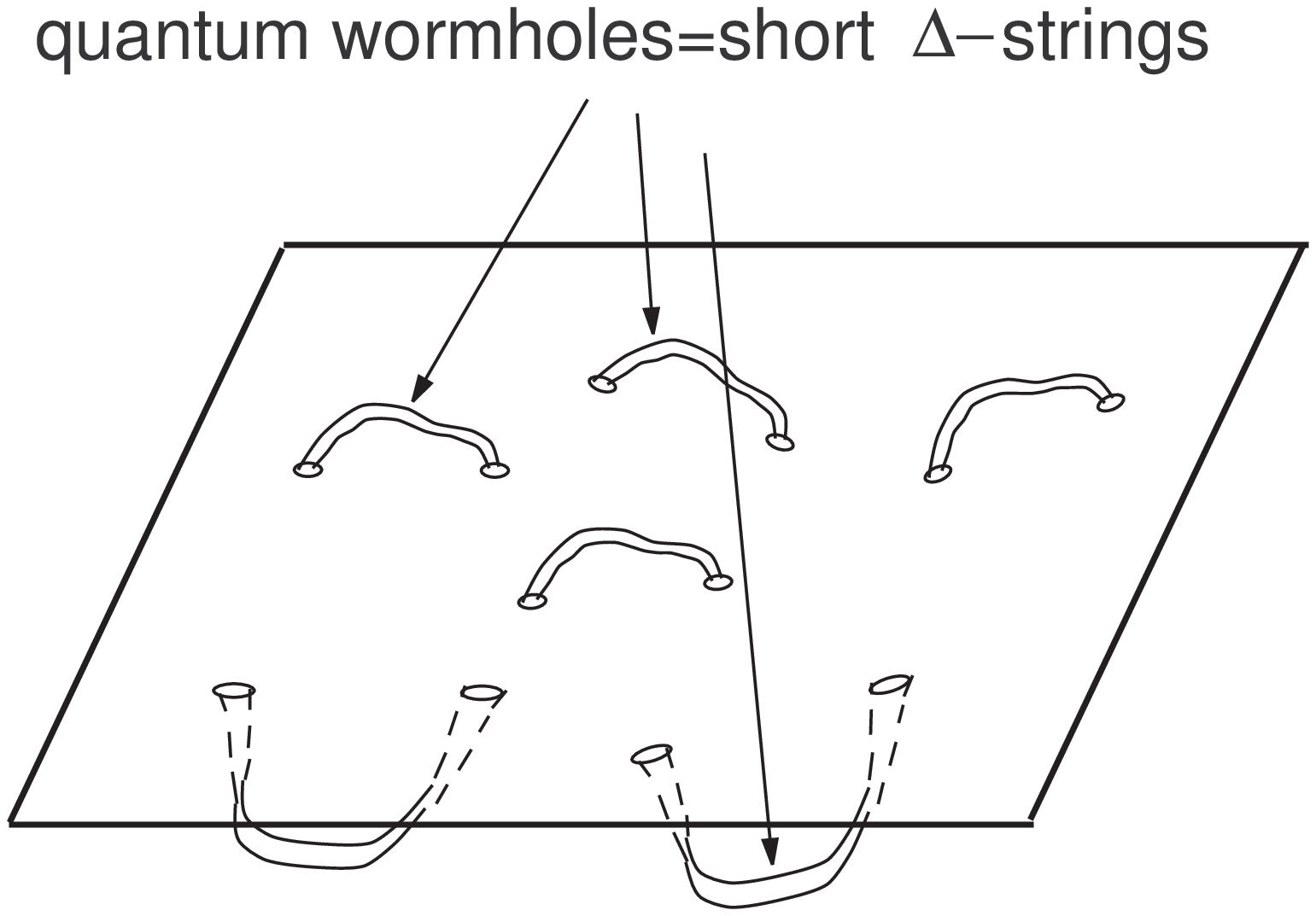}}
\caption{.}
\label{qwormholes}
\end{center}
\end{figure}

\section{Acknowledgment}
I am very grateful to the ISTC grant KR-677 for the financial support.

\end{document}